\documentclass[a4paper,11pt]{article}
\pdfoutput=1
\usepackage{graphicx}
\usepackage{jheppub}
\usepackage{amsmath,amssymb,amsfonts,mathtools,bbold}
\usepackage{braket}
\usepackage[dvipsnames]{xcolor}
\usepackage{tikz}
\usepackage{orcidlink}

\newcommand{\CosmoLattice}{\texttt{${\mathcal C}$osmo${\mathcal L}$attice}}
\newcommand{\MPl}{M_{\mathrm{Pl}}}
\newcommand{\dd}[2][]{\mathinner{\mathrm{d}^{#1}\hspace{-.05em}{#2}}}

\title{Gravitational waves from self-resonance during reheating with a quantum-corrected inflaton potential}

\author[a]{Tomoya Inada\,\orcidlink{0009-0009-4444-9110},}
\emailAdd{inada-tomoya@g.ecc.u-tokyo.ac.jp}

\author[b]{Noel Jonathan Jobu\,\orcidlink{0009-0006-5168-4504},}
\emailAdd{nj702@snu.edu.in}

\author[b]{Kenji Nishiwaki\,\orcidlink{0000-0001-6526-6799},}
\emailAdd{kenji.nishiwaki@snu.edu.in}

\author[a,c]{Toshifumi Noumi\,\orcidlink{0000-0003-0628-8554}}
\emailAdd{tnoumi@g.ecc.u-tokyo.ac.jp}

\author[d]{and Naoki Yamatsu\,\orcidlink{0000-0002-3090-0355}}
\emailAdd{naoki.yamatsu@yukawa.kyoto-u.ac.jp}

\affiliation[a]{Graduate School of Arts and Sciences, The University of Tokyo, Komaba, Meguro-ku, Tokyo 153-8902, Japan}
\affiliation[b]{Department of Physics, School of Natural Sciences,
  Shiv Nadar Institution of Eminence (Deemed to be University),
Tehsil Dadri, Gautam Buddha Nagar, Uttar Pradesh, 201314, India}
\affiliation[c]{RIKEN Center for Interdisciplinary Theoretical and Mathematical Sciences (iTHEMS), RIKEN, Wako 351-0198, Japan}
\affiliation[d]{Yukawa Institute for Theoretical Physics, Kyoto University, Kyoto, Kyoto 606-8502, Japan}

\abstract{
We investigate how localized quantum corrections to the inflaton potential affect preheating dynamics and the resulting stochastic gravitational wave (GW) spectrum.
When these corrections sufficiently suppress the quadratic term of the potential near its minimum, inflaton self-resonance can produce a peaked GW spectrum.
On the other hand, we find that a smooth and enhanced spectrum can appear if the quadratic term acquires a negative coefficient.
As a concrete realization, we analyze the $\alpha$-attractor T-model with a one-loop Coleman--Weinberg correction induced by a heavy scalar and compute the resulting GW spectra using lattice simulations.
The GW signals lie in the ultra-high-frequency regime at frequencies above the kHz range.
These results suggest that GW signals from preheating may probe quantum corrections to the inflaton potential, thereby providing indirect information about the underlying UV physics.

}

\begin{document}

\preprint{RIKEN-iTHEMS-Report-26, YITP-26-51}

\setcounter{tocdepth}{2}
\maketitle
\flushbottom

\section{Introduction}
\label{sec:introduction}

Gravitational waves (GWs) offer a unique and exciting opportunity to directly probe the early universe before recombination.
In contrast to electromagnetic waves, which cannot directly reach us from epochs before recombination, GWs can travel almost freely once they are produced.
Given that the typical energy scale of inflation is far beyond the reach of ground-based particle physics experiments, GWs can also be regarded as a probe of ultra-high-energy physics.
Indeed, various GW production mechanisms have been discussed in the context of inflation and post-inflation dynamics.

Among such sources, (p)reheating~\cite{Dolgov:1989us,Traschen:1990sw,Shtanov:1994ce,Kofman:1994rk,Kofman:1997yn}
has attracted attention as a mechanism for generating a high-frequency stochastic GW background.
See, e.g.,~\cite{Dufaux:2007pt,Allahverdi:2010xz,Amin:2014eta,Lozanov:2019jxc} for reviews.
During this epoch, nonlinear field evolution and resonant amplification
can leave characteristic imprints on the GW spectrum.\footnote{
For recent studies of self-resonance, see, e.g.,~\cite{Jedamzik:2010dq,Child:2013ria,Hertzberg:2014iza,Hertzberg:2014jza,Sfakianakis:2018lzf,Martin:2020fgl,Zhang:2023hjk,del-Corral:2024vcm,Ballesteros:2024eee,Jia:2024fmo}.
For related studies of scalar-induced GWs, see, e.g.,~\cite{Ananda:2006af,Baumann:2007zm,Assadullahi:2009nf,Guzzetti:2016mkm,Domenech:2021ztg,Kohri:2018awv,Franciolini:2021nvv}.
}
More specifically, it was recently shown that local features near the minimum of the inflaton potential can significantly modify the preheating dynamics and the resulting GW spectrum~\cite{Saha:2024lil,Gu:2026ajw}.
Meanwhile, high-frequency GWs in the MHz to GHz range are attracting attention as a new target for future experiments~\cite{Aggarwal:2020olq,Aggarwal:2025noe}.
This raises the possibility of using high-frequency GWs to probe such local features and their underlying ultraviolet (UV) origin.

In this paper, we explore a concrete realization of local features induced by quantum corrections from UV degrees of freedom.
Consider a heavy field coupled to the inflaton $\phi$ through an effective mass $m(\phi)$.
Then, the one-loop Coleman--Weinberg correction to the inflaton potential scales as $\sim m^4(\phi)\ln m^2(\phi)$.
If the effective mass becomes large as the inflaton approaches the potential minimum, while remaining smaller in the inflationary region, the resulting quantum correction is localized near the bottom of the potential.
This provides a simple mechanism for generating localized features that affect reheating dynamics without significantly modifying the inflationary dynamics on cosmic microwave background (CMB) scales.

It is also worth noting that a significant field dependence of the particle spectrum is well motivated from the quantum gravity perspective.
For example, according to the distance conjecture~\cite{Ooguri:2006in}, a super-Planckian excursion in moduli space is typically accompanied by a tower of states whose masses vary exponentially with the canonical field distance.
This suggests that, especially in large-field inflation, the spectrum of heavy degrees of freedom can change significantly between the inflationary region and the potential minimum.
Such a field-dependent spectrum provides another motivation for localized quantum corrections to the inflaton potential.

As an illustrative example of this general mechanism, we study the $\alpha$-attractor T-model~\cite{Kallosh:2013yoa} with an additional heavy scalar coupled to the inflaton~\cite{Kallosh:2016gqp}.
Due to the pole structure of the field-space metric, a simple polynomial dependence of the heavy-field mass on the noncanonical field $\psi$ translates into a localized profile in terms of the canonical inflaton $\phi$. In our benchmark model, the bosonic Coleman--Weinberg correction gives a negative contribution to the quadratic term around the origin.
Based on lattice simulations with \CosmoLattice~\cite{Figueroa:2021yhd} (see also~\cite{Figueroa:2020rrl,Baeza-Ballesteros:2025tme}), we show that a pronounced GW peak can appear through inflaton self-resonance when the quadratic term is sufficiently suppressed and the quartic and higher-order terms control the oscillatory reheating dynamics.
We also show that the peak structure and amplitude depend sensitively on the pole scale of the T-model.

Throughout this paper, we consider a homogeneous and isotropic universe described by the metric
\begin{equation}
  \dd{s}^2 = -\dd{t}^2 + a^2(t)\, \dd{\mathbf{x}}^2\,,
\end{equation}
where $a(t)$ denotes the scale factor.

To make the origin of GWs as transparent as possible, we focus only on self-resonance of the inflaton $\phi$, namely the resonant amplification of the inflaton fluctuations induced by the oscillating homogeneous background.
In a generic reheating scenario, the inflaton can transfer its energy to other fields, commonly called daughter fields, which can affect both the resonance dynamics and the resulting GW signal.
In this paper, however, we neglect such fields and their production.
The same line of analysis can be straightforwardly extended to systems including daughter fields.
We also restrict our discussion to $\mathbb{Z}_2$-symmetric inflaton potentials, $V(-\phi)=V(\phi)$, for simplicity.

\paragraph{Outline.}
The rest of this paper is organized as follows: In Sec.~\ref{sec:resonance_GW}, we review how localized deformations of the inflaton potential can trigger self-resonance during reheating and estimate the characteristic scales of the resulting GW spectrum.
In Sec.~\ref{sec:T_model_qc}, as a concrete realization, we introduce the $\alpha$-attractor T-model with localized one-loop quantum corrections and discuss the parameter region relevant for the subsequent dynamics.
In Sec.~\ref{sec:GW_simulation}, we present {\CosmoLattice} simulations and show how the GW spectra depend on the pole scale and the size of the quantum correction.
In Sec.~\ref{sec:conclusion}, we conclude the paper with a summary and discussion.
In App.~\ref{sec:CMB}, we summarize the observational constraints on the scenario from the current CMB data.
In App.~\ref{sec:CosmoLattice}, we provide technical details of our numerical simulations with {\CosmoLattice}.

\paragraph{Convention.}
We use natural units $c=\hbar=1$.
The reduced Planck mass is defined as $\MPl \coloneq (8 \pi G)^{-1/2} \simeq 2.44 \times 10^{18}\,\mathrm{GeV}$.
Latin indices $i,j,\ldots$ label the three spatial directions, and repeated spatial indices are summed over.

\section{Preheating dynamics in deformed potentials}
\label{sec:resonance_GW}

In this paper, motivated by the reheating scenario proposed in~\cite{Saha:2024lil}, we consider an inflaton potential of the form
\begin{equation}
  V(\phi) = V_0(\phi) + \Delta V(\phi)\,,
  \label{eq:deformed_potential}
\end{equation}
where $V_0(\phi)$ is a slow-roll potential with a minimum at the origin $\phi=0$ and $\Delta V(\phi)$ is a quantum correction localized around the origin.
Such a deformation leaves the inflationary dynamics essentially intact, while modifying the subsequent reheating dynamics.
As shown in~\cite{Saha:2024lil}, this can lead to visible changes in the GW spectrum generated during reheating.

In this section, we summarize why such local deformations can modify the GW spectrum generated during reheating, and under what conditions a sizable GW signal can arise. In particular, we will see that the inflaton undergoes self-resonance during reheating if the quadratic term is subdominant and nonlinear dynamics is relevant around the origin.
We also estimate the characteristic scale of the peak position in the GW spectrum. More detailed spectra will be studied by numerical simulations in Sec.~\ref{sec:GW_simulation} based on a concrete model introduced in the next section.

We first review in Sec.~\ref{subsec:inflaton_fluctuation_resonance} how the local expansion of the inflaton potential controls self-resonance during reheating and determines the characteristic amplified scales.
We then explain in Sec.~\ref{subsec:GW_from_inflaton_fluctuations} how the resulting inflaton inhomogeneities source stochastic GWs.

\subsection{Parametric resonance of inflaton fluctuations}
\label{subsec:inflaton_fluctuation_resonance}

Decompose the inflaton field $\phi(t,\mathbf{x})$ into a homogeneous background $\phi_0(t)$ and an inhomogeneous fluctuation part $\delta\phi(t,\mathbf{x})$ as
\begin{equation}
  \phi(t,\mathbf{x}) = \phi_0(t) + \delta\phi(t,\mathbf{x})\,.
\end{equation}
Then, the background equation of motion and the linearized equation of motion for fluctuations read
\begin{align}
  \ddot{\phi}_0 + 3H\dot{\phi}_0 + V_{,\phi}(\phi_0) &= 0\,,
  \label{eq:background_eom_general} \\
  \delta\ddot{\phi}_{\mathbf{k}} + 3H\delta\dot{\phi}_{\mathbf{k}}
  + \left(\frac{k^2}{a^2}+V_{,\phi\phi}(\phi_0)\right)\delta\phi_{\mathbf{k}} &= 0\,,
  \label{eq:delta_phi_linear_general}
\end{align}
where $H \coloneq \dot{a}/a$ is the Hubble parameter, and we have neglected scalar metric perturbations.
Here, overdots denote derivatives with respect to $t$, and commas in subscripts denote field derivatives, e.g., $V_{,\phi}\coloneq \dd{V}/\dd{\phi}$.
During reheating, $\phi_0$ oscillates around the potential minimum, so that $V_{,\phi\phi}(\phi_0)$ in \eqref{eq:delta_phi_linear_general} acts as a periodic driving term.
If the fluctuation $\delta\phi$ has resonant modes for this driving, the modes become unstable and experience exponential growth.

In general, it is not easy to study the resonance structure analytically.
However, if the potential near the origin is well approximated by a monomial, the corresponding resonance properties are well understood.

First, in the absence of the localized feature $\Delta V(\phi)$, the inflaton potential is generically well approximated by a quadratic form $V(\phi) \sim \phi^2$ near the origin.
The background field then oscillates sinusoidally, with its amplitude damped by Hubble friction.
However, since $V_{,\phi\phi}$ is constant in this case, the fluctuation equation \eqref{eq:delta_phi_linear_general} does not depend on the background field, and hence parametric self-resonance does not occur.

On the other hand, if the localized deformation in \eqref{eq:deformed_potential} cancels out the quadratic term near the origin, the quartic term becomes the leading term and the potential is approximated by
\begin{equation}
  V(\phi) = \frac{\lambda}{4} \phi^4\,.
  \label{eq:quartic_potential}
\end{equation}
In contrast to the quadratic case, $V_{,\phi\phi}(\phi_0)=3\lambda\phi_0^2$ depends on $\phi_0$, so resonant fluctuation modes can appear.
To estimate the scales at which these modes appear, we discuss the resonance structure of the quartic potential in more detail below.

Substituting \eqref{eq:quartic_potential} into \eqref{eq:background_eom_general}, one obtains a nonlinear equation for the background field.
Accordingly, the subsequent oscillation dynamics and resonance structure depend nontrivially on the initial conditions.
We define $t_*$ as the time of the first turning point of the damped oscillation, at which
\begin{equation}
\dot{\phi}_0(t_*)=0\,.
\end{equation}
We assume without loss of generality that inflation happens in the regime $\phi<0$ and the first turning point is in the positive region $\phi_0(t_*)>0$.
The initial amplitude of the damped oscillation is then given by
\begin{equation}
\phi_* \coloneq \phi_0(t_*)\,.
\label{eq:phi_star_definition}
\end{equation}
In the following, we solve for the oscillatory background evolution starting from $t=t_*$ with these initial conditions.

To study the resulting resonance structure more explicitly, it is convenient to introduce the dimensionless conformal variables~\cite{Greene:1997fu,Figueroa:2016wxr}:
\begin{equation}
  \mathbf{y} \coloneq \sqrt{\lambda} \, \phi_*\,\mathbf{x}\,, \qquad
  z \coloneq \sqrt{\lambda}\,\phi_*\,(\eta-\eta_*)\,, \qquad
  \dd\eta \coloneq \frac{\dd t}{a(t)}\,,
\end{equation}
\begin{equation}
  \varphi_0(z) \coloneq a(\eta) \frac{\phi_0(\eta)}{\phi_*}\,, \qquad
  \delta\varphi(z,\mathbf{y}) \coloneq a(\eta) \frac{\delta\phi(\eta,\mathbf{x})}{\phi_*}\,.
\end{equation}
Here $\eta_*$ is the conformal time corresponding to the physical time $t_*$ and we choose the normalization of the scale factor as $a(\eta_*) = 1$ for simplicity.
In terms of these variables, the equations for the background field and fluctuation become
\begin{align}
  \varphi_0'' + \varphi_0^3 &= \frac{a''}{a}\varphi_0\,, \\
  \delta\varphi_{\tilde{\mathbf{k}}}'' + \left(\tilde{k}^2 + 3\varphi_0^2 \right)\delta\varphi_{\tilde{\mathbf{k}}} &=  \frac{a''}{a} \delta\varphi_{\tilde{\mathbf{k}}}\,,
  \label{eq:conformal_delta_phi_eom}
\end{align}
where primes denote derivatives with respect to $z$, $\tilde{\mathbf{k}} \coloneq \mathbf{k}/(\sqrt{\lambda}\,\phi_*)$, and $\tilde{k} \coloneq |\tilde{\mathbf{k}}|$.
For the quartic potential, the background expansion is of radiation-domination type, so $a''=0$ and the right-hand sides vanish.
The background field then admits an oscillatory solution in terms of the Jacobi elliptic cosine,
\begin{equation}
  \varphi_0(z) = \operatorname{cn}{\left(z,\frac{1}{2}\right)}\,,
\end{equation}
with the initial conditions $\varphi_0(0)=1$ and $\varphi_0'(0)=0$.

With this background solution, the fluctuation equation \eqref{eq:conformal_delta_phi_eom} reduces to the Lam\'e equation with resonance parameter $q=3$.
Its resonance structure is well known, and the dominant resonance band is peaked around~\cite{Greene:1997fu,Figueroa:2016wxr}
\begin{equation}
  \tilde{k}_\mathrm{p} = \frac{k_\mathrm{p}}{\sqrt{\lambda}\, \phi_*} \simeq 1.27\,.
  \label{eq:quartic_resonance_mode}
\end{equation}

Thus, the resonance structure can be estimated from the behavior of the potential near the origin.
Notice, however, that the above discussion is self-consistent only if the field value $\phi_*$ lies within the regime of validity of the expansion around the origin.
If $\phi_*$ lies outside that regime, the potential can no longer be approximated by a single monomial, and the resonance structure becomes more complicated, typically amplifying modes over a broad range rather than at a peak scale.

\subsection{Production of gravitational waves}
\label{subsec:GW_from_inflaton_fluctuations}

We next briefly review how the self-resonant inflaton fluctuations source GWs.
Write the spatial metric as
\begin{equation}
  \dd s^2 = -\dd t^2 + a^2(t)\left(\delta_{ij}+h_{ij}\right)\dd x^i \dd x^j\,,
\end{equation}
where $h_{ij}$ satisfies the transverse-traceless (TT) conditions $\partial_i h_{ij}=0$ and $h_{ii}=0$. Then, the equation of motion in the presence of a TT source $\Pi_{ij}^{\mathrm{TT}}$ is
\begin{equation}
  \ddot{h}_{ij}+3H\dot{h}_{ij}-\frac{\nabla^2}{a^2}h_{ij}
  = \frac{2}{\MPl^2}\Pi_{ij}^{\mathrm{TT}}\,.
\end{equation}
The associated GW energy density is
\begin{equation}
  \rho_{\mathrm{GW}}(t)=\frac{\MPl^2}{4}\left\langle \dot{h}_{ij}\dot{h}_{ij}\right\rangle\,,
\end{equation}
where $\langle\cdots\rangle$ denotes a spatial average, and the GW energy density spectrum per logarithmic momentum interval is then defined as
\begin{equation}
  \Omega_{\mathrm{GW}}(k,t)\coloneq \frac{1}{\rho_{\mathrm{c}}(t)}\frac{\dd \rho_{\mathrm{GW}}(t)}{\dd \ln k}
  \,,
  \qquad
  \rho_{\mathrm{c}}(t)=3\MPl^2 H^2(t)\,.
\end{equation}

During preheating, the amplified inflaton inhomogeneities generate anisotropic stress and thereby source stochastic GWs.
For the inflaton fluctuation, the TT source is
\begin{equation}
  \Pi_{ij}^{\mathrm{TT}} = \frac{1}{a^2}\left[\partial_i\delta\phi\,\partial_j\delta\phi\right]^{\mathrm{TT}}\,.
\end{equation}
The corresponding GW spectrum can be expressed in conformal time $\eta$ as~\cite{Figueroa:2017vfa}
\begin{equation}
  \Omega_{\mathrm{GW}}(k,\eta) = \frac{k^3}{48\pi^4\MPl^4 H^2}
  \int \dd p\, \dd\theta\, p^6 \sin^5\theta
  \left(
    \left|I_{(c)}(k,p,\theta,\eta)\right|^2
    + \left|I_{(s)}(k,p,\theta,\eta)\right|^2
  \right)\,,
  \label{eq:GW_spectrum_delta_phi}
\end{equation}
where
\begin{align}
  I_{(c)}(k,p,\theta,\eta)
  &\coloneq \int_{\eta_i}^{\eta} \dd\eta'\, a(\eta') \cos(k\eta')\,
  \delta\phi_{\mathbf{k}-\mathbf{p}}(\eta')\,\delta\phi_{\mathbf{p}}(\eta')\,,
  \label{eq:GW_int_cos}\\
  I_{(s)}(k,p,\theta,\eta)
  &\coloneq \int_{\eta_i}^{\eta} \dd\eta'\, a(\eta') \sin(k\eta')\,
  \delta\phi_{\mathbf{k}-\mathbf{p}}(\eta')\,\delta\phi_{\mathbf{p}}(\eta')\,.
  \label{eq:GW_int_sin}
\end{align}
Here, $\eta_i$ denotes the initial time at which the source term starts to be integrated, and $\theta$ denotes the angle between $\mathbf{k}$ and $\mathbf{p}$.

Since the fluctuation spectrum peaks around $k_{\mathrm{p}}$ in \eqref{eq:quartic_resonance_mode}, one can infer from the integrals in \eqref{eq:GW_int_cos} and \eqref{eq:GW_int_sin} that the resulting GWs are produced mainly in the range $0 \le k \le 2k_{\mathrm{p}}$.
However, obtaining the exact GW spectrum requires carrying out the time and momentum integrals, and it is difficult to determine the peak position analytically.
Here we restrict ourselves to the qualitative expectation that the GW spectrum peaks at a scale comparable to $k_{\mathrm{p}}$.

\section{T-model and quantum corrections}
\label{sec:T_model_qc}

In this section, we show that quantum corrections to the inflaton potential can generate a localized deformation in \eqref{eq:deformed_potential}.
As implied by the discussion in the previous section, such corrections can change whether self-resonance occurs and hence whether GWs are produced efficiently.
From this perspective, GW observations may provide an indirect probe of quantum corrections to the inflaton potential.

As a concrete example of the general mechanism stated in Sec.~\ref{sec:introduction}, we consider the $\alpha$-attractor T-model and discuss the corresponding quantum correction.\footnote{
Cosmological attractors introduced in 2013 include conformal attractors~\cite{Kallosh:2013hoa}, $\xi$-attractors~\cite{Kallosh:2013tua}, and $\alpha$-attractors~\cite{Ferrara:2013rsa,Kallosh:2013yoa}; see also~\cite{Ellis:2013nxa,Kallosh:2014rga,Roest:2015qya,Linde:2015uga,Scalisi:2015qga,Carrasco:2015pla}.
Generalizations include P-models with singular derivatives of the potential~\cite{Kallosh:2022feu} and S-models with singular potentials~\cite{Linde:2017pwt,Linde:2018hmx,Kallosh:2025sji}.
Supergravity embeddings of $\alpha$-attractors are discussed in~\cite{Cecotti:2014ipa,Kallosh:2015zsa,Carrasco:2015uma,Kallosh:2025dac}, string embeddings in~\cite{Cicoli:2008gp,Kallosh:2017wku,Cicoli:2024bxw} (see also~\cite{Kallosh:2024ymt}), and associated axion fields in~\cite{Kallosh:2010xz,Kallosh:2017wnt,Achucarro:2017ing,Linde:2018hmx,Kallosh:2021fvz,Carrasco:2025rud}.  
}
After briefly reviewing the tree-level T-model in Sec.~\ref{subsec:T_model}, we introduce the one-loop correction generated by a heavy scalar in Sec.~\ref{subsec:quantum_correction}.
These ingredients define the effective inflaton potential used in the lattice simulations performed in the next section.

\subsection{T-model inflation}
\label{subsec:T_model}

T-model inflation is motivated by a hyperbolic internal geometry, and the kinetic term of the inflaton $\psi$ has a pole structure:
\begin{equation}
  \mathcal{L} = \frac{\MPl^2}{2}R
  - \frac{(\partial\psi)^2}{2 \left(1-\psi^2/M^2\right)^2}
  -V_0(\psi)\,,
\end{equation}
where $M$ specifies the location of the pole and $V_0(\psi)$ is a potential.
In terms of the canonically normalized field $\phi$ defined through
\begin{equation}
  \psi = M \tanh{\left(\frac{\phi}{M}\right)}\,,
\end{equation}
the Lagrangian reads
\begin{equation}
  \mathcal{L} = \frac{\MPl^2}{2}R
  - \frac{1}{2}(\partial\phi)^2
  - V_0{\left(M \tanh{\left(\frac{\phi}{M}\right)}\right)}\,.
\end{equation}
The simplest and most widely used choice of the inflaton potential is
\begin{equation}
  V_0(\psi) = \frac{\Lambda^4}{2}\left(\frac{\psi}{M}\right)^2
  = \frac{\Lambda^4}{2} \tanh^2{\left(\frac{\phi}{M}\right)}\,,
  \label{eq:tree_potential}
\end{equation}
where we assumed a $\mathbb{Z}_2$ symmetry of $\psi$ and truncated higher-order terms in $\psi$, assuming that they are suppressed by a UV scale.
We make similar assumptions throughout the paper.
For the inflaton potential~\eqref{eq:tree_potential}, the scalar tilt $n_\mathrm{s}$ and the tensor-to-scalar ratio $r$ read~\cite{Kallosh:2013yoa}
\begin{equation}
  \label{eq:T_model_ns_r}
  n_\mathrm{s}\simeq 1-\frac{2}{N_e},
  \qquad
  r\simeq \frac{2}{N_e^2}\left(\frac{M}{\MPl}\right)^2\,,
\end{equation}
where $N_e$ is the $e$-folding number.
This gives $n_\mathrm{s}\simeq 0.960$--$0.967$ for $N_e \simeq 50$--$60$ and also naturally gives a small $r$ for $M \lesssim \MPl$, compatible with CMB observations by the Planck collaboration~\cite{Planck:2018jri,BICEP:2021xfz}.
See App.~\ref{sec:CMB} for more details on the comparison, including more recent experiments such as ACT and SPT.

\subsection{One-loop quantum corrections in the T-model}
\label{subsec:quantum_correction}

As a concrete realization of the localized feature, we consider a heavy scalar $\sigma$ coupled to the inflaton through a $\psi$-dependent effective mass $m(\psi)$:\footnote{
In general, there may exist more fields that induce quantum corrections to the inflaton potential.
While we focus on the simplest setup with a single heavy scalar, our analysis remains qualitatively the same as long as the dominant source of the quantum correction is a bosonic field.
On the other hand, if the dominant source is a fermion, the quantum correction has the opposite overall sign because of the Grassmann-odd nature.
As a result, the type of inflaton self-resonance discussed in Sec.~\ref{sec:resonance_GW} does not occur, and hence no pronounced peak is expected in the GW spectrum.}

\begin{equation}
  \mathcal{L}_{\sigma} = -\frac{1}{2}(\partial\sigma)^2 - \frac{1}{2}m^2(\psi)\sigma^2,\qquad
  m^2(\psi)= m_0^2 + a_2 \psi^2\,.
\end{equation}
Here we assume that $m^2(\psi)$ respects the $\mathbb{Z}_2$ symmetry $\psi\to-\psi$ and is truncated at quadratic order in $\psi$, where $a_2$ is a constant coefficient.
For simplicity, we suppress the self-interaction of $\sigma$.

We can integrate out $\sigma$ if, throughout the relevant range of $\psi$, its effective mass is sufficiently large compared with the Hubble scale $H_{\mathrm{inf}}$ during inflation and the characteristic oscillation scale $\omega_{\mathrm{reh}}$ during reheating:
\begin{equation}
  m(\psi) \gg H_{\mathrm{inf}}\,,\qquad
  m(\psi) \gg \omega_{\mathrm{reh}}
  \sim \sqrt{|V_{0,\phi\phi}(0)|} = \frac{\Lambda^2}{M}\,.
\end{equation}
In this case, the effective potential including the one-loop Coleman--Weinberg correction \cite{Coleman:1973jx} (see also~\cite{Weinberg:1973am}) is\footnote{
  Note that the effective potential is one-loop exact in the present setup, since $\sigma$ has no self-interaction.
}
\begin{equation}
  V_{\mathrm{eff}}(\psi) = V_0(\psi) + \frac{m^4(\psi)}{64\pi^2} \ln{\left(\frac{m^2(\psi)}{m_0^2}\right)}\,,
\end{equation}
where we regard the tree-level potential $V_0(\psi)$ as a renormalized one, and assume that terms beyond quadratic order in $\psi$ remain suppressed at this scale. We also require that the vacuum energy vanishes at the potential minimum so that the post-inflationary universe has a vanishing cosmological constant. If the potential minimum is at $\psi=\psi_{\min} \neq 0$, we further include a constant counterterm such that $V_{\mathrm{eff}}(\psi_{\min})=0$, as discussed later.
Since $\psi=M\tanh(\phi/M)$ becomes constant at large $|\phi|$, this quantum correction does not significantly affect the inflationary dynamics, while it can modify the reheating dynamics through deformations of the potential shape around the origin.

\begin{figure}[t]
  \centering
  \includegraphics[width=0.7\textwidth]{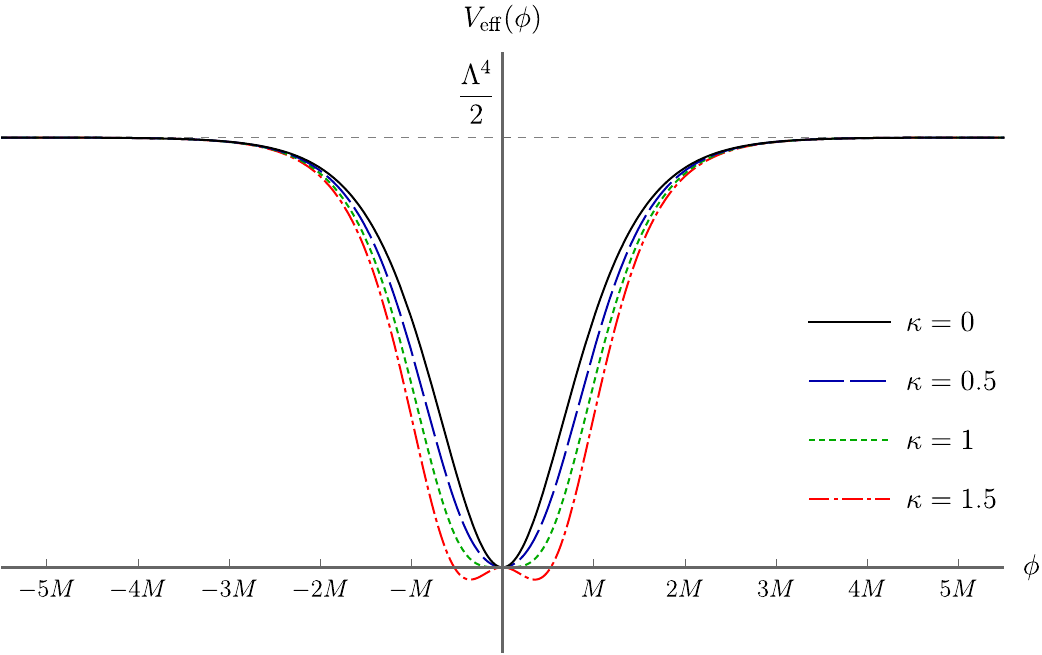}
  \caption{Effective potential for representative values of $\kappa$.
  Increasing $\kappa$ enhances the local deformation around the origin, while leaving the large-field plateau almost unchanged.
  Note that, for $\kappa>1$, the potential minima are shifted away from the origin.
  Therefore, an additive constant should be included so that the minimum is lifted to zero.
  In this figure, however, this constant shift is omitted to make the comparison of the potential shapes clearer.}
  \label{fig:potential_plot}
\end{figure}

In the following, we examine how the quantum correction changes the shape of the inflaton potential.
  As a simple benchmark example, we first consider the specific coupling
  \begin{equation}
    a_2 = - \frac{m_0^2}{M^2}\,.
    \label{eq:a2_benchmark}
  \end{equation}
This choice is useful because the quantum correction vanishes at large $|\phi|$.
With this choice, the effective potential as a function of the canonical field $\phi$ is
\begin{equation}
  \label{eq:Veff_full}
  V_{\mathrm{eff}}(\phi)=\frac{\Lambda^4}{2} \left[
    \tanh^2{\left(\frac{\phi}{M}\right)}
    - 2 \kappa \cosh^{-4}{\left(\frac{\phi}{M}\right)}
    \ln{\left(\cosh{\left(\frac{\phi}{M}\right)}\right)}
  \right]\,,
\end{equation}
where we parametrized the magnitude of the quantum correction using the dimensionless parameter,
\begin{equation}
  \kappa \coloneq \frac{m_0^4}{32\pi^2 \Lambda^4}\,.
  \label{eq:kappa-definition}
\end{equation}
Fig.~\ref{fig:potential_plot} shows how the effective potential changes as $\kappa$ is varied.\footnote{
Notice here that $\kappa$ is not a parameter of the perturbative expansion but is introduced for convenience to quantify the one-loop correction coefficient relative to the tree-level coefficient.}
As $\kappa$ increases, the potential is more strongly deformed near the origin, while the large-field behavior relevant for inflation remains almost unchanged. See App.~\ref{sec:CMB} for compatibility of the quantum-corrected inflaton potential with the CMB constraints.

For a qualitative discussion of the reheating dynamics, it is useful to expand the effective
potential around the origin:
\begin{equation}
  V_{\mathrm{eff}}(\phi)=\frac{\Lambda^4}{2}\left[
    (1-\kappa)\left(\frac{\phi}{M}\right)^2
    + \frac{-4+13\kappa}{6}\left(\frac{\phi}{M}\right)^4
    + \mathcal{O}\!\left(\left(\frac{\phi}{M}\right)^6\right)
  \right].
  \label{eq:Veff_origin_expansion}
\end{equation}
Note that this expansion works only when $|\phi|$ is sufficiently small compared with $M$ and breaks down when $|\phi|\gtrsim M$.
Now the role of $\kappa$ is transparent from this expression:
For $\kappa<1$, the quadratic term dominates near the origin.
At $\kappa = 1$, the quadratic coefficient vanishes, so that the quartic term becomes dominant near the origin when $\kappa\simeq 1$.
For $\kappa>1$, the quadratic coefficient becomes negative, and the origin turns into an unstable point.
As a result, the inflaton oscillation pattern can change qualitatively in these three different regimes.
As discussed in Sec.~\ref{sec:resonance_GW}, changing the dominant term around the origin affects the resonance structure and, consequently, the resulting GW spectrum.

We emphasize that this cancellation does not rely on fine-tuning of the coupling constant to the benchmark value in \eqref{eq:a2_benchmark}.
  For $-m_0^2/M^2 \le a_2 <0$, the same cancellation of the quadratic coefficient can occur after an appropriate choice of the relative size of the quantum correction, or equivalently of $m_0/\Lambda$.
  In such cases, the quantum correction generally gives a finite constant contribution in the large-field region.
  However, since the overall height of the potential is ultimately fixed by choosing $\Lambda$ to reproduce the observed scalar amplitude, this constant contribution has little impact on the inflationary dynamics~\cite{Kallosh:2016gqp}.
  In contrast, for $a_2 \ge 0$, the one-loop correction has the same sign as the tree-level quadratic term near the origin, and such a cancellation does not occur.
Also, for $a_2 < -m_0^2/M^2$, the effective mass squared $m^2(\psi)$ becomes negative for sufficiently large field values, so the heavy field develops a tachyonic instability and the effective potential above is no longer applicable.\footnote{
    Strictly speaking, even at the value in \eqref{eq:a2_benchmark}, the validity of the effective potential is subtle in the large-field limit, where the effective mass squared $m^2(\psi)$ vanishes.
    In practice, however, inflation probes only a finite field range and does not reach this limit exactly.
    Therefore, if $m(\psi)$ is sufficiently large over the field range relevant for inflation, the effective potential remains under control.
    In addition, this choice is only a benchmark within the broader parameter range $-m_0^2/M^2 \le a_2 < 0$, and this issue does not arise away from the boundary value.
    Therefore, we do not discuss it further in this paper.}

  In the next section, focusing on the coupling choice in \eqref{eq:a2_benchmark}, we use numerical simulations to examine the GW spectra in the different regimes of $\kappa$.
In the simulations, we use the exact effective potential rather than its expansion around the origin.

\section{Gravitational wave spectra from lattice simulations}
\label{sec:GW_simulation}

In this section, we study the GW spectra obtained from lattice simulations based on the effective potential in \eqref{eq:Veff_full}.
While the potential has three parameters $\Lambda,M,\kappa$, the overall scale $\Lambda$ of the potential can be fixed by the scalar amplitude $A_\mathrm{s}$, once we specify the $e$-folding number $N_e$ (see also App.~\ref{sec:CMB}). In our analysis, we choose $N_e=60$ for illustration. The potential is then parameterized by the pole scale $M$ and the deformation parameter $\kappa$. In the following subsections, we study three representative values of the pole scale, $M=\MPl$, $M=0.1\,\MPl$, and $M=0.01\,\MPl$, while varying $\kappa$ in each case.

To simulate the reheating dynamics and compute the resulting GW spectra, we use {\CosmoLattice}~\cite{Figueroa:2021yhd}, a public lattice code for fields in an expanding spacetime.
We focus on self-resonance by neglecting interactions with daughter fields, although such interactions can be included in the same framework.
We set the lattice parameters and initial conditions so that the relevant dynamics is captured within the simulated range.
The spatial lattice contains at least $N^3=256^3$ points, with $N^3=384^3$ used when a wider spectral range is required.
For each parameter set, the IR cutoff $k_{\mathrm{IR}}$ is fixed so that the relevant spectral features lie well between $k_{\mathrm{IR}}$ and the UV cutoff $k_{\mathrm{UV}} \simeq \pi/{\dd{x}}$, with the box size then given by $L=2\pi/k_{\mathrm{IR}}$ and $\dd{x}=L/N$.
We choose the time step to satisfy $\dd{t} \lesssim {\dd{x}}/3$ for the stability of the time evolution.
Further details of the initial conditions and simulation parameters are provided in App.~\ref{sec:CosmoLattice}.

In all plots presented in this section, we show the GW spectra as they would be observed today.
To obtain them, we take into account the redshift between the end of the simulation, $t_{\mathrm{f}}$, and the present time, $t_{\mathrm{0}}$.
The GW frequency and energy density spectrum at the present time are given by~\cite{Figueroa:2017vfa,Cosme:2022htl}
\begin{align}
  f_{\mathrm{GW},0} &\simeq 4 \times 10^{10} \, \epsilon_{\mathrm{f}}^{1/4}
  \frac{k}{a_{\mathrm{f}} H_{\mathrm{f}}}
  \left(\frac{H_{\mathrm{f}}}{\MPl}\right)^{1/2} \, \mathrm{Hz} \, , \\
  h_0^2 \Omega_{\mathrm{GW},0} &\simeq 1.6 \times 10^{-5} \, \epsilon_{\mathrm{f}} \, \Omega_{\mathrm{GW},\mathrm{f}} \, .
\end{align}
Here, $H_{\mathrm{f}}$, $a_{\mathrm{f}}$, and $\Omega_{\mathrm{GW},\mathrm{f}}$ are the Hubble parameter, scale factor, and GW energy density spectrum at $t_{\mathrm{f}}$, all read directly from the simulation.
In addition, the factor $\epsilon_{\mathrm{f}}$ parametrizes the redshift between $t_{\mathrm{f}}$ and the onset of the radiation-dominated era, $t_{\mathrm{RD}}$, as
\begin{equation}
  \epsilon_{\mathrm{f}} \coloneq \left(\frac{a_{\mathrm{f}}}{a_{\mathrm{RD}}}\right)^{1-3\bar{w}}\, ,
\end{equation}
where $\bar{w}$ is the effective equation of state during the intermediate era.
When the potential is approximately quartic, the homogeneous oscillating field has an effective equation of state $\bar{w} \simeq 1/3$, and $\epsilon_{\mathrm{f}}$ is therefore close to unity.
Since our main focus is the quartic-like resonant regime and its nearby parameter region, we set $\epsilon_{\mathrm{f}}=1$ for simplicity.\footnote{
  A more precise determination of $a_{\mathrm{f}}/a_{\mathrm{RD}}$ and $\bar{w}$ requires following the thermalization process, which is beyond the scope of this paper.
  For the quartic-like regime considered here, especially around $\kappa\simeq 1$, we expect $\epsilon_{\mathrm{f}}\simeq 1$ to be a good approximation.
}

In the following, we present the GW spectra and discuss their physical interpretations for $M=\MPl$, $M=0.1\,\MPl$, and $M=0.01\,\MPl$, in turn.

\begin{figure}[t]
  \centering
  \includegraphics[width=0.6\textwidth]{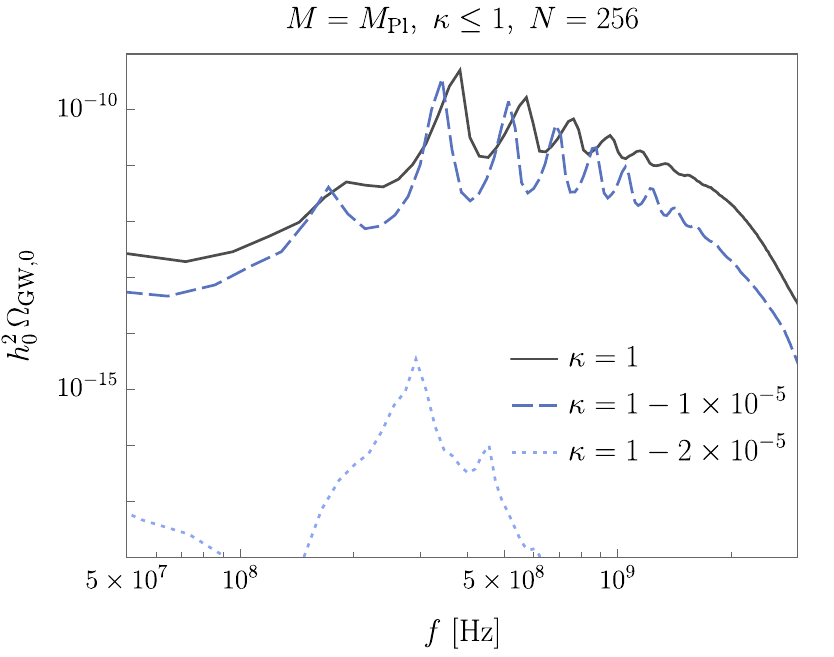}
  \caption{GW spectra at $M=\MPl$ for several values of $\kappa\leq 1$, with $N=256$.
  The characteristic multi-peak structure observed at $\kappa=1$ is strongly suppressed when $\kappa$ is lowered from unity by an amount of $\mathcal{O}(10^{-5})$.}
  \label{fig:gw_M1}
\end{figure}

\subsection{Large pole scale: \texorpdfstring{$M = \MPl$}{M = MPl}}
\label{subsec:sim_M1}

In light of the tensor-to-scalar ratio in \eqref{eq:T_model_ns_r}, the observationally allowed upper end of the pole scale is of order $M \sim \MPl$.
With this in mind, we first consider the case $M=\MPl$. 

\begin{figure}[t]
  \centering
  \begin{minipage}[t]{0.49\textwidth}
    \centering
    \includegraphics[width=\textwidth]{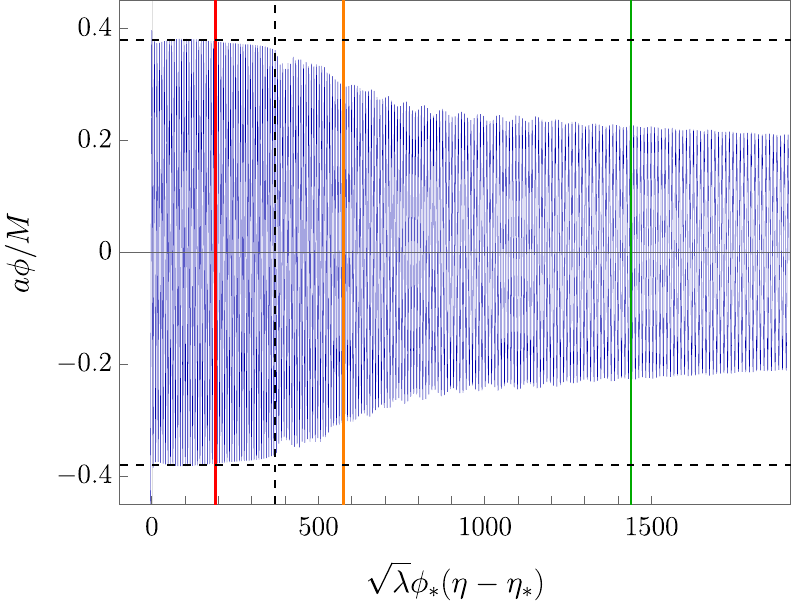}
  \end{minipage}
  \begin{minipage}[t]{0.49\textwidth}
    \centering
    \includegraphics[width=\textwidth]{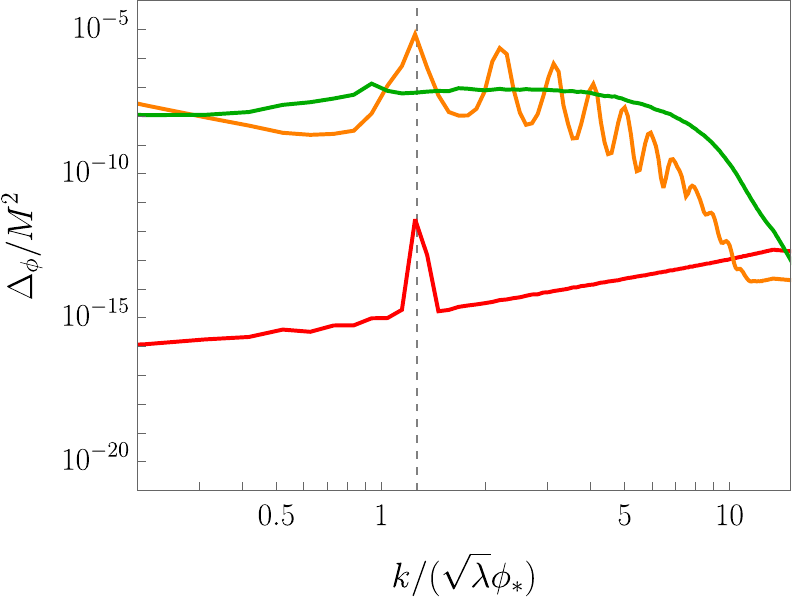}
  \end{minipage}
  \caption{
    Left: Evolution of the spatially averaged rescaled inflaton field, $a\phi/M$, for $M=\MPl$ and $\kappa=1$. The horizontal dashed lines indicate the approximately constant oscillation envelope during the resonant stage, $|a\phi| = \phi_* \simeq 0.38M$.
    The vertical dashed line marks the onset of nonlinear effects in the fluctuations, which we identify with the time when the rescaled amplitude ceases to remain approximately constant, estimated as
    $\sqrt{\lambda}\phi_*(\eta-\eta_*) \simeq 370$.
    Right: Dimensionless power spectra of inflaton fluctuations at representative times, with colors corresponding to the vertical lines in the left panel.
    The dominant resonance initially appears near $\tilde{k} \simeq 1.27$, and nonlinear rescattering later generates additional peaks.
  }
  \label{fig:M1_dynamics}
\end{figure}

\paragraph{Symmetric phase.}

We begin with the region $\kappa \le 1$, which we call the symmetric phase because the $\mathbb{Z}_2$ symmetry remains unbroken at the potential minimum.
The resulting GW spectra are shown in Fig.~\ref{fig:gw_M1}.

Inspecting the critical case $\kappa=1$, one finds that multiple peaks appear in the GW spectrum.
This behavior can be understood from the expansion in \eqref{eq:Veff_origin_expansion}.
At $\kappa=1$, the quadratic term vanishes and the effective potential around the origin is well approximated by a quartic monomial,
\begin{equation}
  V_{\mathrm{eff}}(\phi) \simeq \frac{\lambda}{4}\phi^4,
  \qquad
  \lambda \coloneq \frac{3\Lambda^4}{M^4}\,.
  \label{eq:M1_quartic}
\end{equation}
To check compatibility with this approximation, Fig.~\ref{fig:M1_dynamics} shows the evolution of the rescaled homogeneous mode and the dimensionless power spectra of the inflaton,
\begin{equation}
  \braket{\phi(\mathbf{k}) \phi(\mathbf{k}')}
  \eqcolon \frac{\Delta_\phi(k)}{2k^3} (2\pi)^5 \delta^{(3)}(\mathbf{k}+\mathbf{k}')\,,
\end{equation}
at representative stages of the simulation.
First, shortly after the onset of resonance, the spectrum (red line) exhibits a peak at $\tilde{k} \simeq 1.27$.
This is consistent with the value of $\tilde{k}_{\mathrm{p}}$ in \eqref{eq:quartic_resonance_mode} and shows agreement with the approximation in \eqref{eq:M1_quartic}.

On the other hand, at later times, multiple peaks emerge in the spectrum (orange line).
This can be understood as a consequence of the breakdown of the linear approximation once the inflaton fluctuations become sufficiently large: nonlinear rescattering then excites modes around integer multiples of $k_{\mathrm{p}}$.
Indeed, at the corresponding times the rescaled amplitude starts to decrease, indicating that the fluctuations have grown large enough for backreaction on the homogeneous background to become important.\footnote{For a detailed discussion of backreaction and the associated peaks in GW spectra, see~\cite{Figueroa:2016wxr,Figueroa:2017vfa}.}
Eventually, the rescaled amplitude becomes approximately constant, and the peaks in the power spectrum become smoother as the rescattering process saturates (green line).

The peak positions in the GW spectrum are not expected to coincide exactly with those in the scalar spectrum, because the source is quadratic in $\delta \phi$ and involves a convolution over momenta.
Nevertheless, if the scalar spectrum contains pronounced peaks around $k \simeq n k_{\mathrm{p}}$, the convolution enhances GW production at momenta corresponding to vector sums of two such modes.
This explains why the GW spectrum retains multiple peaks, even though the individual peak positions are shifted relative to the scalar spectrum.

Spectra with such a peak structure are also observed when $\kappa$ is smaller than unity by an amount of order $10^{-5}$.
However, for smaller $\kappa$, GW production is abruptly suppressed.
This is because the quadratic term becomes dominant during oscillations, thereby suppressing the quartic resonance.
Quantitatively, to obtain a spectrum with peaks, the following condition on the field value $\phi$ has to be satisfied for a while during the oscillatory phase:
\begin{equation}
  (1-\kappa)\left(\frac{\phi}{M}\right)^2 \ll \frac{-4+13\kappa}{6}\left(\frac{\phi}{M}\right)^4 \, .
  \label{eq:kappa_condition_M1}
\end{equation}
In particular, for the peak amplitude to be comparable to that obtained for $\kappa=1$, this condition must remain satisfied for a period comparable to the duration of the resonance in that case.
Estimating the end of the resonance stage in which the linear approximation remains valid and a distinct peak is present from Fig.~\ref{fig:M1_dynamics} as $\sqrt{\lambda}\,\phi_* (\eta-\eta_*) \sim 370$ and substituting the corresponding field amplitude $|\phi| \simeq 8.6 \times 10^{-3} M$, one finds that the condition \eqref{eq:kappa_condition_M1} breaks down at $1 - \kappa \simeq 10^{-4}$, which is consistent with the numerical result in Fig.~\ref{fig:gw_M1}.

\paragraph{Broken phase.}

Next, we turn to the region $\kappa > 1$.
As shown in Fig.~\ref{fig:potential_plot}, in this regime the effective potential in \eqref{eq:Veff_full} develops degenerate minima with negative vacuum energy at $\phi=\pm\phi_{\mathrm{min}}\neq 0$, so that the $\mathbb{Z}_2$ symmetry is spontaneously broken.
We therefore add a constant counterterm to lift the minimum to zero.
Also, for this reason, we call the regime $\kappa>1$ the broken phase.

\begin{figure}[t]
  \centering
  \begin{minipage}[t]{0.49\textwidth}
    \centering
    \includegraphics[width=\textwidth]{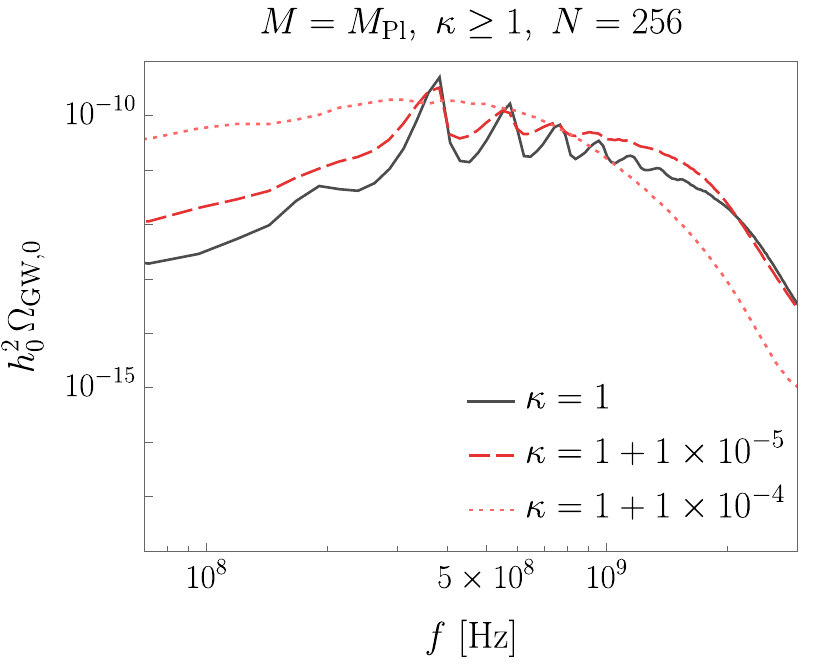}
  \end{minipage}
  \begin{minipage}[t]{0.49\textwidth}
    \centering
    \includegraphics[width=\textwidth]{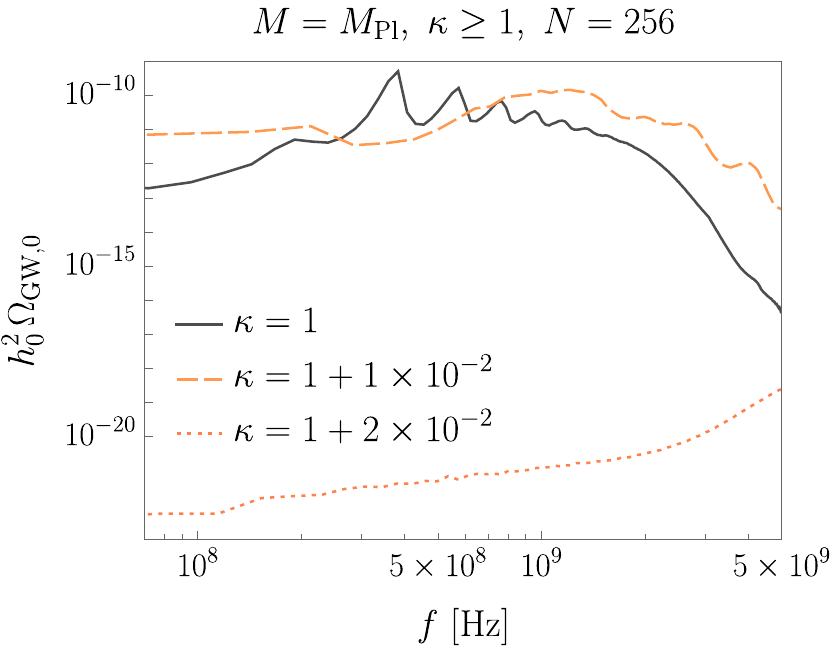}
  \end{minipage}
  \caption{GW spectra at $M=\MPl$ for several values of $\kappa\geq 1$, with $N=256$.
  The left panel shows that, when $\kappa$ is raised above unity by an amount of $\mathcal{O}(10^{-4})$, the characteristic multi-peak structure observed at $\kappa=1$ disappears and is replaced by a smooth spectrum.
  The right panel shows that this smooth spectrum is also strongly suppressed when $\kappa$ is further increased by an amount of $\mathcal{O}(10^{-2})$.  We also note that, at the boundary value $\kappa=1+10^{-2}$, a distinctive spectral shape is observed: a step-like feature appears in the high-frequency region above $10^{9}\,\mathrm{Hz}$.
  }
  \label{fig:gw_M1_gt1}
\end{figure}

The corresponding GW spectra obtained from the lattice simulations are shown in Fig.~\ref{fig:gw_M1_gt1}.
Again, multiple peaks are visible when $\kappa$ is close to unity by an amount of order $10^{-5}$.
However, once $\kappa$ exceeds unity by an amount of order $10^{-4}$, the peak structure disappears, and we instead find a smooth spectrum.

This behavior can be attributed to the bump that develops in the potential around the origin, which drives the system away from the resonance characteristic of a quartic potential.
In this regime, the inflaton oscillations can no longer be described by a monomial approximation to the potential around the origin.
Consequently, unlike in the $\kappa<1$ case, the second derivative of the potential $V_{,\phi\phi}(\phi_0)$ in \eqref{eq:delta_phi_linear_general} has a nontrivial effect on the evolution of the fluctuations.
We speculate that the resulting more complicated resonance structure excites a broad range of modes rather than a few sharply peaked bands, thereby giving rise to the smooth GW spectrum.

Another possible interpretation of the smooth GW spectrum is tachyonic preheating~\cite{Kofman:1997yn,Felder:2000hj}.
When the potential contains a region of negative curvature $V_{\mathrm{eff},\phi\phi}<0$, the effective mass squared of the fluctuations in \eqref{eq:delta_phi_linear_general} can become negative as the background field passes through this region. Modes satisfying $k^2/a^2<|V_{\mathrm{eff},\phi\phi}|$ then undergo exponential growth. In contrast to quartic self-resonance, which preferentially amplifies modes within characteristic resonance bands, this tachyonic instability excites a broad range of IR modes below a characteristic momentum scale.
The resulting broadband enhancement provides a natural explanation for the smooth, rather than multi-peaked, GW spectrum.\footnote{We note that~\cite{Gu:2026ajw} discusses the possibility that parametric resonance and tachyonic instability can coexist within a single preheating scenario.
Since a detailed analysis of this possibility is beyond the scope of this paper, we do not pursue it further here.}

As can also be seen from the right panel of Fig.~\ref{fig:gw_M1_gt1}, such smooth spectra disappear for $\kappa = 1 + \mathcal{O}(10^{-2})$.
This can be understood as follows: as the central bump becomes more pronounced, the inflaton oscillates predominantly around $\phi_{\mathrm{min}}$ rather than around the origin, so that the reheating dynamics can be approximated by the quadratic term around the minimum, rendering the resonance inefficient.

Finally, we briefly comment on a peculiar step-like feature that appears near the threshold value $\kappa = 1 + 10^{-2}$ in the high-frequency region, $f \gtrsim 10^9\,\mathrm{Hz}$.
Since this feature appears in a region somewhat below the UV cutoff, it appears to be a physical signal rather than a UV artifact.
However, it depends sensitively on the infrared cutoff: if $k_{\mathrm{IR}}$ is chosen too large, the overall amplitude of the spectrum is strongly suppressed.\footnote{See also Fig.~\ref{fig:conv_IR} in App.~\ref{sec:CosmoLattice}, where this case is presented as an illustrative example of the generic dependence of GW spectra on the IR cutoff.
}
Therefore, resolving the detailed shape and origin of this feature would require simulations with a rather large lattice size $N$.
In addition, since the parameter region where it appears is very narrow, we leave a detailed investigation of its physical origin for future work.

\subsection{Intermediate pole scale: \texorpdfstring{$M=0.1\,\MPl$}{M = 0.1 MPl}}
\label{subsec:sim_M0.1}

Next, we decrease the pole scale to $M=0.1\,\MPl$ and perform the same analysis.
This allows us to examine how changing the value of $M$ affects the GW spectrum.

\paragraph{Symmetric phase.}

\begin{figure}[t]
  \centering
  \includegraphics[width=0.6\textwidth]{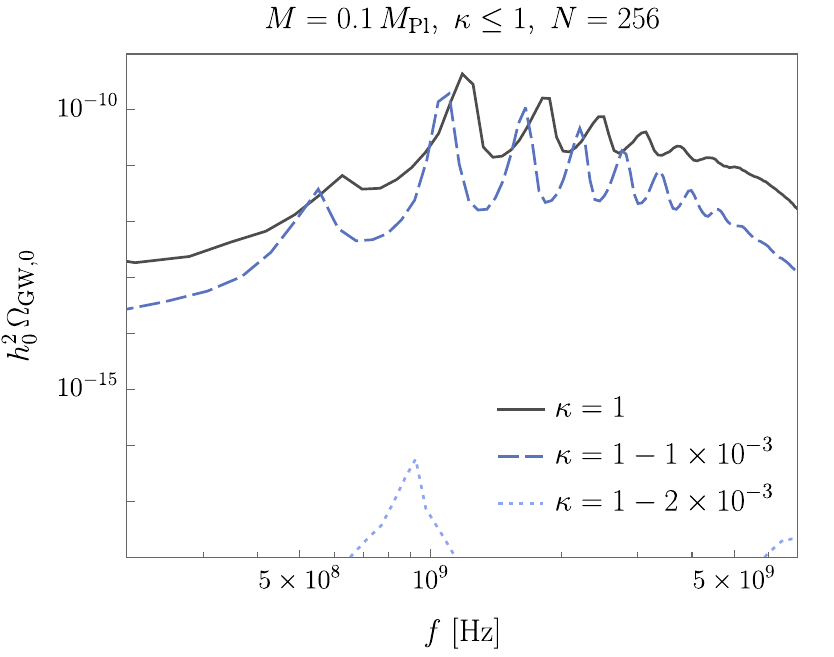}
  \caption{GW spectra at $M=0.1\,\MPl$ for several values of $\kappa\leq 1$, with $N=256$.
  The characteristic multi-peak structure observed at $\kappa=1$ is strongly suppressed when $\kappa$ is lowered from unity by an amount of $\mathcal{O}(10^{-3})$. Compared with the $M=\MPl$ case, resonant GW production persists over a much wider range of $\kappa$.}
  \label{fig:gw_M0.1}
\end{figure}

The GW spectra in this setup are shown in Fig.~\ref{fig:gw_M0.1}.
The simulation at $\kappa=1$ again shows pronounced peaks.
However, compared to the $M=\MPl$ case, these peaks survive over a wider range, down to $\kappa = 1 - \mathcal{O}(10^{-3})$.

This is because decreasing the pole scale also lowers the height of the potential.
As a result, the background-field velocity is less efficiently damped by Hubble friction, allowing the field to travel farther before reaching its first turning point and thus increasing $\phi_*$.
Then the ratio $\phi_*/M$, which is relevant in the condition~\eqref{eq:kappa_condition_M1}, is larger than in the previous case, so the quartic term becomes more dominant.
Consequently, we see that the peaked GW signal 
survives in a wider window of $\kappa$ around 
unity.\footnote{For a similar discussion, see~\cite{Lozanov:2019ylm}.}

Even so, the allowed window in $\kappa$ is still rather narrow and requires a certain amount of fine tuning.
One might expect that taking $M$ even smaller would lead to a wider window.
However, as we will see in Sec.~\ref{subsec:sim_M0.01}, naively decreasing $M$ too much changes the resonance structure itself.

\paragraph{Broken phase.}

\begin{figure}[t]
  \centering
  \begin{minipage}[t]{0.49\textwidth}
    \centering
    \includegraphics[width=\textwidth]{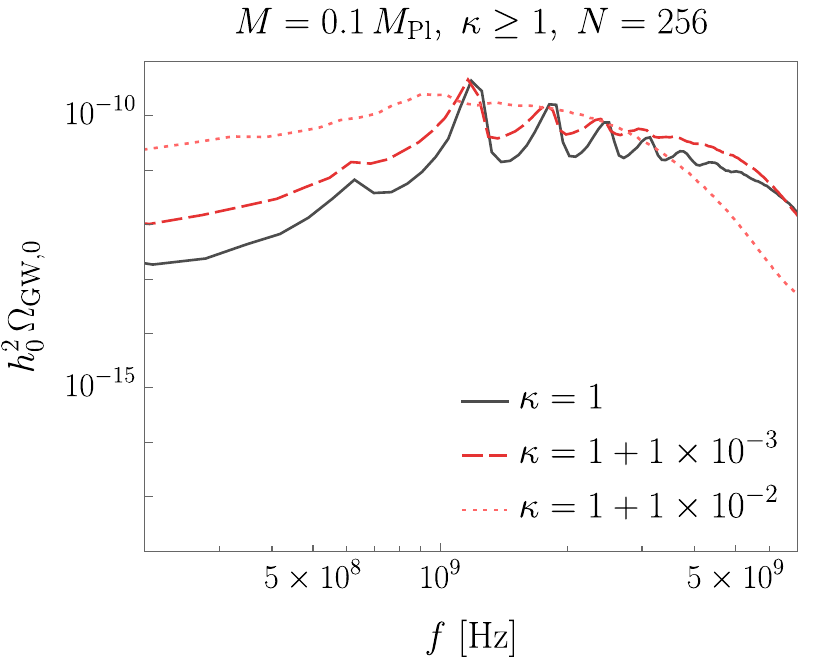}
  \end{minipage}
  \begin{minipage}[t]{0.49\textwidth}
    \centering
    \includegraphics[width=\textwidth]{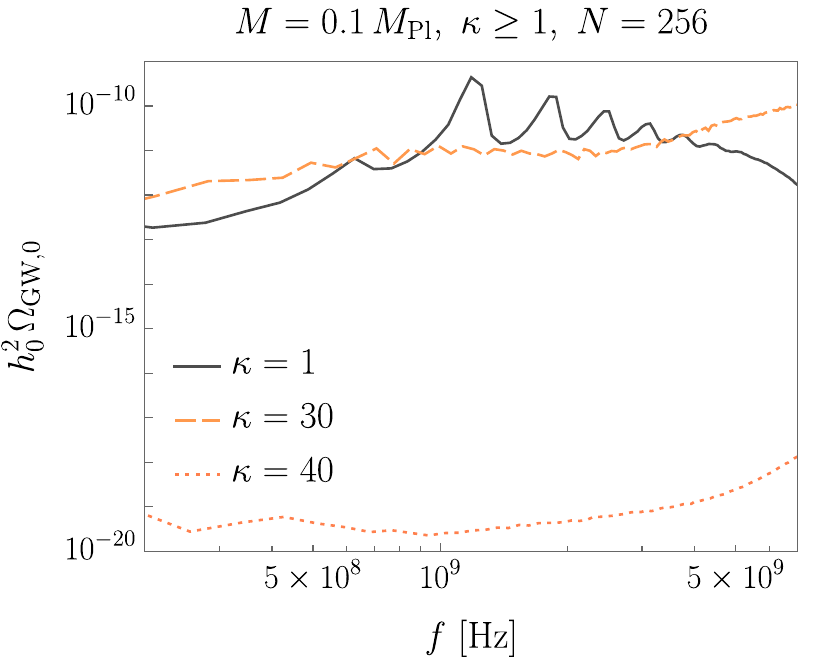}
  \end{minipage}
  \caption{GW spectra at $M=0.1\,\MPl$ for several values of $\kappa\geq 1$, with $N=256$.
  The left panel shows that, when $\kappa$ is raised above unity by an amount of $\mathcal{O}(10^{-2})$, the characteristic multi-peak structure observed at $\kappa=1$ disappears and is replaced by a smooth spectrum.
  The right panel shows that this smooth spectrum is also strongly suppressed when $\kappa$ is further increased by an amount of $\mathcal{O}(10)$. Compared with the symmetric phase, the range of $\kappa$ over which sizable GW signals are produced is much wider.}
  \label{fig:gw_M0p1_gt1}
\end{figure}

For $\kappa>1$, the GW spectra shown in Fig.~\ref{fig:gw_M0p1_gt1} exhibit behavior similar to that in the $M=\MPl$ case.
The peak structure is maintained up to $\kappa = 1 + \mathcal{O}(10^{-3})$, whereas for larger values the spectrum becomes flattened up to $\kappa = \mathcal{O}(10)$.

As in the symmetric phase, the parameter region exhibiting a peaked spectrum is considerably wider than in the $M=\MPl$ case.
Moreover, when the region yielding a nearly smooth spectrum is also included, large-amplitude GW signals are obtained over a remarkably broad parameter range.
This suggests that the enhancement does not require severe fine-tuning and may provide a realistic target for future detection.

\subsection{Smaller pole scale: \texorpdfstring{$M=0.01\,\MPl$}{M = 0.01 MPl}}
\label{subsec:sim_M0.01}

\begin{figure}[t]
  \centering
  \includegraphics[width=0.6\textwidth]{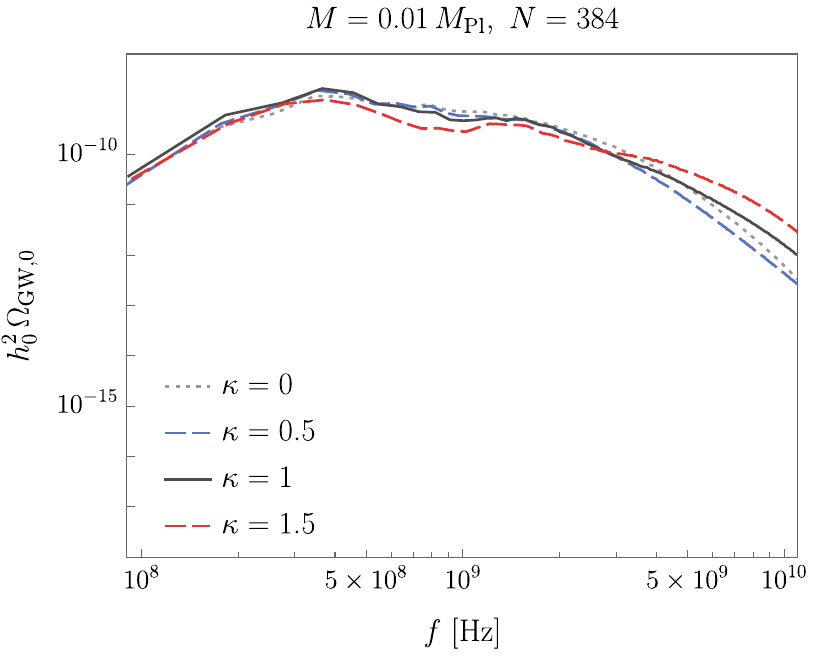}
  \caption{GW spectra at $M=0.01\,\MPl$ for several values of $\kappa$, with $N=384$.
  At this pole scale, the resulting GW spectra are almost independent of $\kappa$, and hence of the detailed shape of the potential near its minimum.
  In all cases, nearly identical spectra characterized by a smooth broadband enhancement are obtained.}
  \label{fig:gw_M0.01}
\end{figure}

From the observation in the previous subsection, one might expect that the peak-producing window in $\kappa$ can be arbitrarily broadened as $M$ decreases.
However, the simulation results in Fig.~\ref{fig:gw_M0.01} show that this naive expectation fails.
In this regime, we have $\phi_* \gtrsim M$ and so the intuition based on the small $\phi$ approximation does not apply.
The effects of higher-order terms are summarized as follows.
First, the GW spectrum is no longer controlled only by the locally quartic part of the potential.
In fact, previous studies of preheating in $\alpha$-attractor models found that even the tree-level potential can already produce an overall uplifted GW spectrum in this regime~\cite{Li:2020gw,Krajewski:2022gw}.
Accordingly, as shown in Fig.~\ref{fig:gw_M0.01}, GWs are generated even for $\kappa=0$, namely already in the tree-level T-model.
Second, once the whole spectrum is lifted rather than showing a clean isolated peak, the integrated GW abundance may approach the standard Big Bang Nucleosynthesis bound~\cite{Yeh:2022heq} on extra radiation.
The small-$M$ regime is therefore not simply a more favorable version of the $\kappa\simeq1$ mechanism.

\paragraph{Artificial initial conditions.}

\begin{figure}[t]
  \centering
  \includegraphics[width=0.6\textwidth]{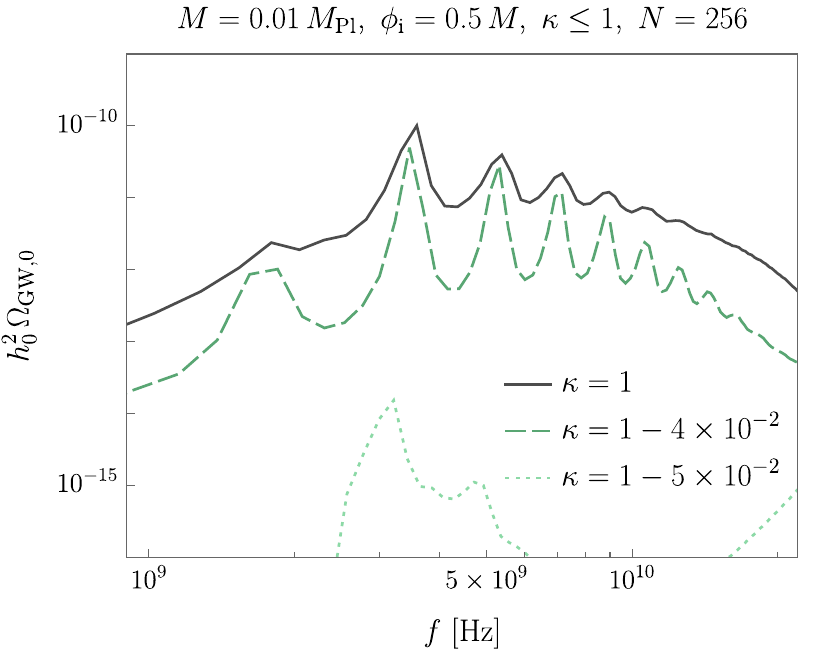}
  \caption{GW spectra at $M=0.01\,\MPl$ for several values of $\kappa \leq 1$, with $N=256$ and the initial field value $\phi_{\mathrm{i}} = - 0.5\,M$. 
  Although artificial in the minimal setup, this initial condition may qualitatively mimic additional dissipative effects that delay the onset of inflaton oscillations to smaller field amplitudes.
  For this choice, the characteristic multi-peak structure observed at $\kappa = 1$ is strongly suppressed when $\kappa$ is lowered from unity by an amount of $\mathcal{O}(10^{-2})$.
}
  \label{fig:gw_smallinit}
\end{figure}

While the standard initial conditions at $M=0.01\,\MPl$ do not lead to a clean isolated peak, such a peak can be recovered if one artificially starts the evolution from a sufficiently small initial field value.\footnote{
  We note that the choice of initial conditions did not pose a problem in the previous sections because the simulations were started before the onset of nonlinear resonance.
  In contrast, if one artificially chooses the starting point inside a field region where resonance should already have begun, as in the present case, the result depends strongly on the initial conditions.
}
As a representative example, Fig.~\ref{fig:gw_smallinit} shows the GW spectra from simulations with the initial conditions $\phi_{\mathrm{i}}=-0.5M$ and $\dot{\phi}_{\mathrm{i}}=0$.
In this case, the peak structure again appears around $\kappa=1$, and the corresponding range extends more widely to $\kappa = 1 - \mathcal{O}(10^{-2})$.

Although this choice is artificial within the minimal setup, it may qualitatively mimic a situation in which interactions with other fields or an additional background energy component enhance the effective damping and prolong the friction-dominated regime.
In particular, it is reminiscent of dissipative dynamics such as warm inflation, in which the inflaton equation of motion contains an additional friction term characterized by a dissipation coefficient $\Gamma$.
Roughly speaking, such a situation can be viewed as one in which the Hubble-friction scale $H(t)$ is replaced by an effective damping scale $H(t) + \Gamma(t)$, up to factors of order unity. 
This enhanced friction slows the inflaton as it rolls down the potential, thereby reducing the value of $\phi_*$.

From this perspective, the recovery of the multi-peak structure for a smaller initial amplitude suggests that the signal may persist in reheating histories more involved than the minimal single-field evolution considered here.
However, additional dissipative effects could themselves modify the inflaton dynamics and the resulting observables in a nontrivial way.
This interpretation should therefore be regarded as qualitative rather than as a self-consistent realization of such a scenario.

\subsection{Summary of parameter dependence}
\label{subsec:summary}

\begin{figure}[t]
  \centering
  \begin{minipage}[t]{0.49\textwidth}
    \centering
    \includegraphics[width=\textwidth]{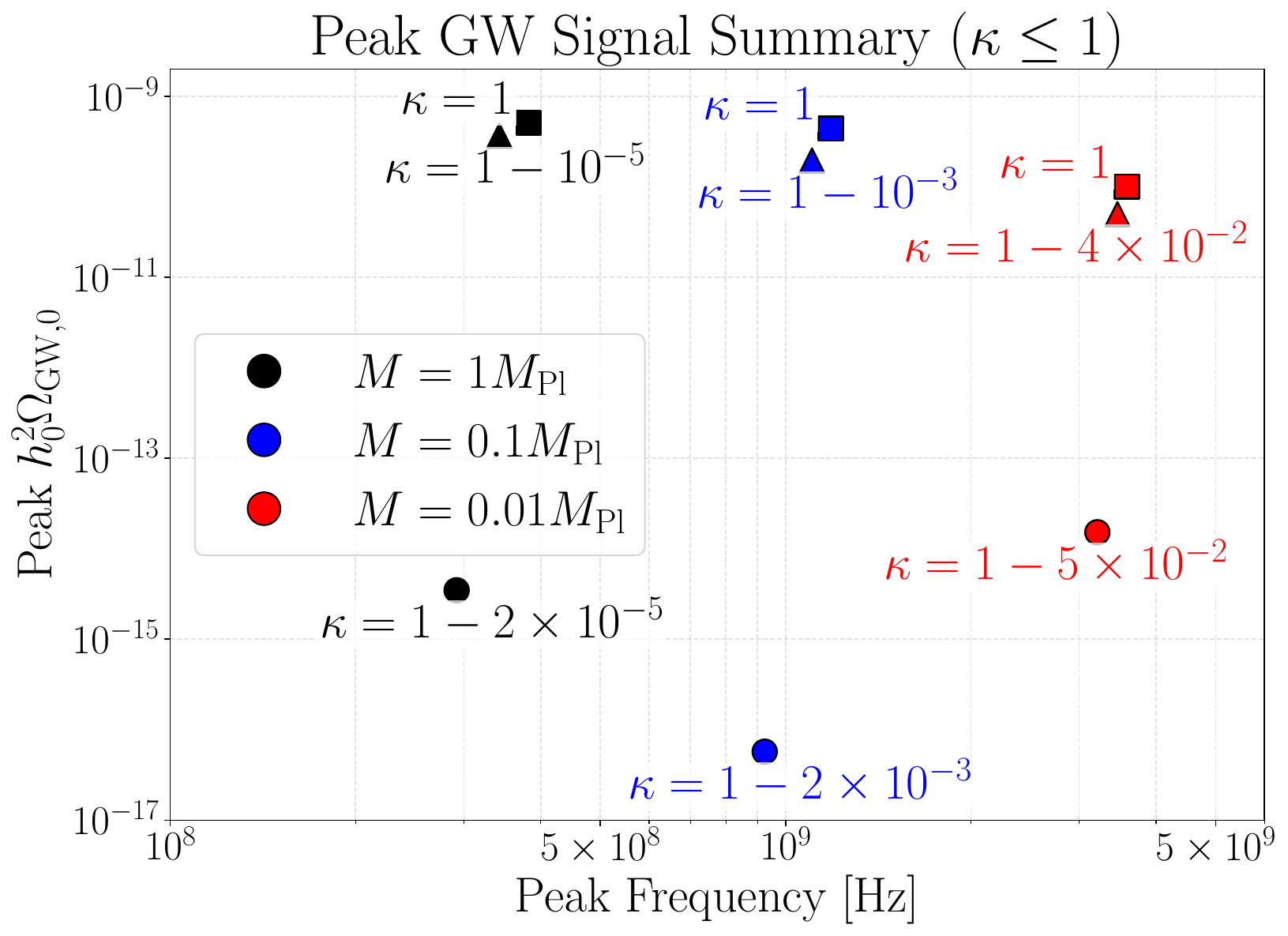}
  \end{minipage}
  \begin{minipage}[t]{0.5\textwidth}
    \centering
    \includegraphics[width=\textwidth]{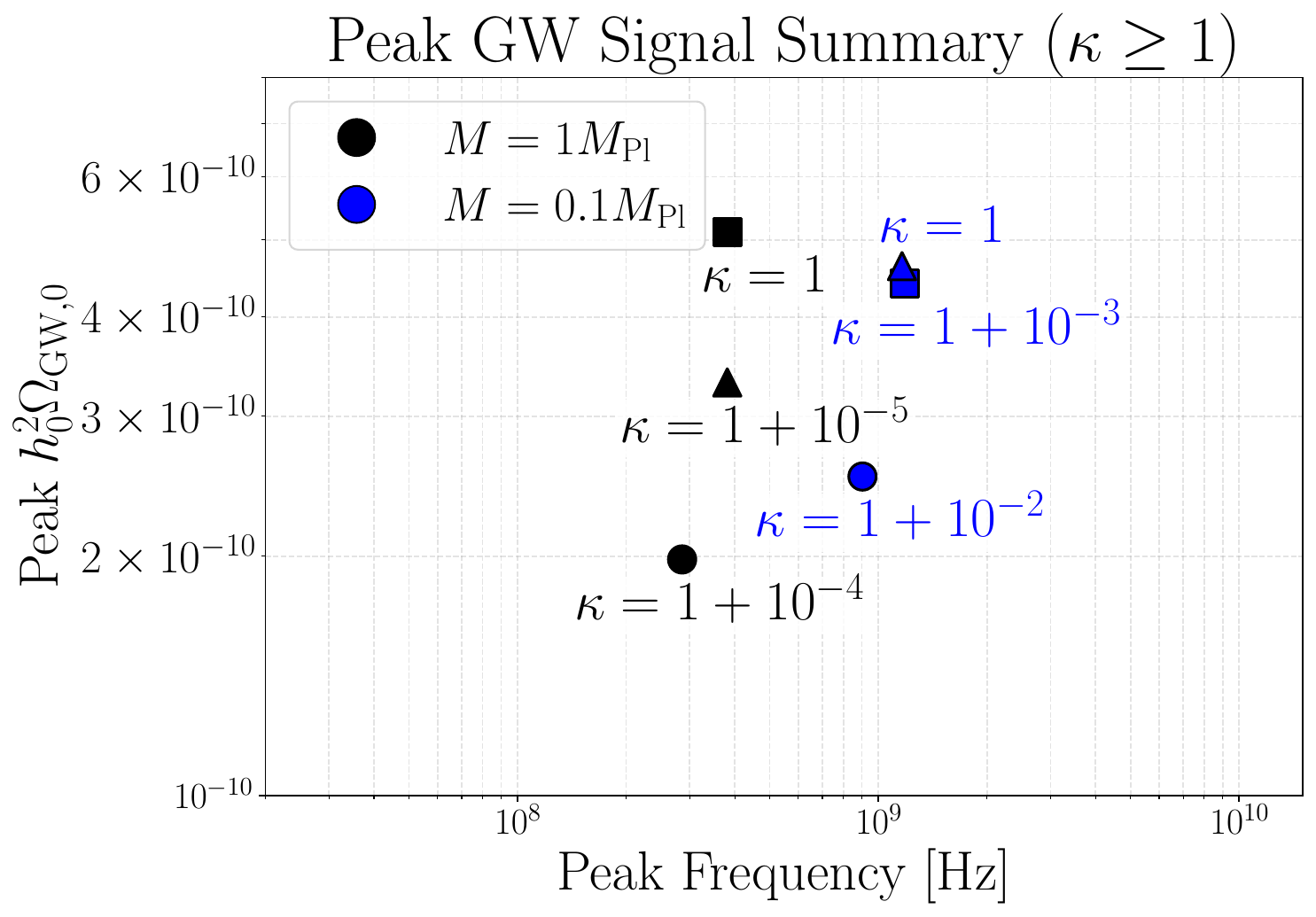}
  \end{minipage}
  \caption{Summary plot of the peak GW energy density $h_0^2\Omega_{\mathrm{GW},0}$ and
      peak frequency $f_{\mathrm{peak}}$ extracted from all lattice simulations.
      Left: results for $\kappa \le 1$.
      Right: results for $\kappa \ge 1$.
      In both panels, the color encodes the pole scale:
      black ($M = \MPl$), blue ($M = 0.1\,\MPl$), and red ($M = 0.01\,\MPl$).
      Within each color group, the marker shape encodes the relative
  magnitude of the deformation. Note: The red points in the left panel correspond to the artificial initial conditions shown in Fig.~\ref{fig:gw_smallinit}.
  }
  \label{fig:summary}
\end{figure}

The spectra discussed in the preceding subsections are summarized in Fig.~\ref{fig:summary} by plotting, for each spectrum, the maximum present-day GW energy density $h_0^2\Omega_{\mathrm{GW},0}$ against the frequency at which it is attained.
For spectra exhibiting multiple peaks, we use the highest peak.
This comparison reveals several characteristic trends.

\paragraph{Dependence on the pole scale $M$.}

Fig.~\ref{fig:summary} (left) reveals a clear dependence of the peak GW signal on the pole mass scale $M$. We may infer that decreasing $M$ shifts the peak frequency to higher values.
This trend can be understood from the fact that a smaller pole scale implies a larger effective quartic $\lambda$ as seen in~\eqref{eq:M1_quartic}.\footnote{
  More precisely, $\lambda$ also depends on the overall potential scale $\Lambda$.
  Although $\Lambda$ also decreases as $M$ is lowered, since this parameter is determined by the scalar amplitude $A_\mathrm{s}$, it is not simply proportional to $M$.
  For the representative numerical values of $\Lambda$, see App.~\ref{sec:CosmoLattice}.}
The larger coupling increases the characteristic resonance momentum, $k_{\mathrm{p}} \propto \lambda^{1/2}\phi_*$, which in turn leads to a higher peak frequency observed today, following the arguments presented in Sec.~\ref{subsec:sim_M1}.
It is also clear that the peak amplitude in all three cases of $M/\MPl = 1, 0.1, 0.01$ is of the order of $10^{-10}$, suggesting that the peak amplitude in the resonant regime is insensitive to the pole scale.

\paragraph{Sensitivity to $\kappa$.}

We also observe that, for $\kappa \le 1$, the system is very sensitive to the deformation parameter $\kappa$, as seen in the left panel of Fig.~\ref{fig:summary}.
At the highest value of $M$, reducing $\kappa$ below unity by as little as
$1 - \kappa \sim 10^{-5}$ suppresses the peak by an
order of magnitude, and by $1 - \kappa \sim 2 \times 10^{-5}$
the signal has dropped to $h_0^2\Omega_{\mathrm{GW},0} \sim 10^{-15}$, five orders of magnitude below the $\kappa = 1$ value.
A similar trend is observed for smaller pole scales, but the range of $\kappa$ over which a pronounced peak is generated in the spectrum is relatively wider.
This broadening is consistent with the analytic estimate in
Sec.~\ref{subsec:sim_M1}: the condition for quartic dominance
during resonance scales as~\eqref{eq:kappa_condition_M1}.

\paragraph{Behavior for $\kappa \ge 1$.}

The right panel of Fig.~\ref{fig:summary} shows the $M = \MPl$ and $M=0.1\,\MPl$ results for $\kappa \ge 1$.
In contrast to the $\kappa \le 1$ case, the peak amplitude decreases
more gradually as $\kappa$ moves away from unity, and the spectrum
becomes smooth rather than collapsing abruptly.
This is because, for $\kappa>1$, the center of oscillation shifts from the origin to $\phi_{\mathrm{min}}\neq 0$, where the potential is no longer well approximated by a purely quartic form.
Consequently, the oscillations exhibit a weaker resonance structure, as explained in Sec.~\ref{subsec:sim_M1}.

\section{Conclusion}
\label{sec:conclusion}

In this paper, we have investigated how localized deformations of the inflaton potential affect the dynamics of preheating and the resulting stochastic gravitational wave (GW) spectrum.
Motivated by the general form $V(\phi)=V_0(\phi)+\Delta V(\phi)$, we have shown that even a localized modification around the origin can significantly alter the resonance structure of inflaton fluctuations.

The key observation is that the efficiency of self-resonance is controlled by the leading term in the potential around the origin at the onset of oscillations.
In particular, when the quadratic term is suppressed and the quartic term becomes dominant, the inflaton fluctuations exhibit parametric resonance with a characteristic scale, leading to a peaked GW spectrum.
On the other hand, when the quadratic term dominates or when the potential cannot be approximated by a single monomial, the resonance becomes inefficient or broad, resulting in a suppressed or smooth GW spectrum.

As a concrete realization of this idea, we have considered the $\alpha$-attractor T-model with a one-loop Coleman--Weinberg correction induced by a heavy scalar field.
The resulting effective potential is characterized by a deformation parameter $\kappa$, which controls the relative importance of the quadratic and quartic terms near the origin.
Using lattice simulations, we have demonstrated that a pronounced peak in the GW spectrum, $h_0^2 \Omega_{\mathrm{GW},0} \sim 10^{-10}$, appears only in a narrow region around $\kappa \simeq 1$, where the quartic term dominates during the oscillatory stage.
Also, the behavior differs significantly between the symmetric phase ($\kappa < 1$) and the broken phase ($\kappa > 1$), separated by the critical point ($\kappa = 1$).

We have also explored the dependence on the pole scale $M$.
For $M \sim \MPl$, the peaked GW signal arises only in a highly tuned region of $\kappa$.
For smaller values of $M$, the quartic-dominated regime can persist for a longer period, slightly broadening the parameter window.
However, when $M$ becomes sufficiently small, the expansion around the origin breaks down and higher-order terms of the potential become important, leading to a qualitatively different GW spectrum.
The fact that self-resonance occurs over a broad parameter region near $\kappa \gtrsim 1$ when $M$ is relatively small compared with $\MPl$ is expected to play an important role in the phenomenological tests of our scenario.

These results suggest that GW observations from preheating may provide a novel probe of quantum corrections to the inflaton potential and may offer indirect access to heavy degrees of freedom in the UV theory.
Although the ultra-high-frequency (UHF) GWs obtained in this work lie at frequencies far above the bands targeted by current and proposed laser-interferometric GW detectors~\cite{LIGOScientific:2014pky,VIRGO:2014yos,Aso:2013eba,KAGRA:2018plz,
2017arXiv170200786A,Seto:2001qf,Harry:2006fi,
Hild:2010id,Punturo:2010zz,Reitze:2019iox}, these signals provide a concrete target for future dedicated detectors in this band~\cite{Aggarwal:2025noe}.
In particular, several UHF search strategies have been proposed~\cite{Berlin:2021txa,Domcke:2022rgu,Herman:2022fau,Ito:2022rxn,Ito:2023fcr,Capdevilla:2024cby}.

Several limitations of the present analysis should be kept in mind.
We have neglected interactions with daughter fields and focused solely on self-resonance.
In addition, we have assumed that the heavy field $\sigma$ has no self-interactions, so that the one-loop approximation remains valid.
It is also assumed that the system reaches radiation domination efficiently.
Including these effects may modify the quantitative predictions and deserves further investigation.

It would be interesting to extend the present analysis in several directions.
In particular, incorporating couplings to additional fields, going beyond the one-loop approximation, and exploring more general classes of inflaton potentials would help clarify the robustness of the mechanism.
Furthermore, a more detailed study of the resulting GW signals, including their frequency range and detectability, would be important for connecting this scenario with future GW observations.

Finally, in this paper, we have considered only the simplest case of the T-model; however, within the context of $\alpha$-attractors, singular models (S-models)~\cite{Linde:2017pwt,Linde:2018hmx,Kallosh:2025sji}, which modify the behavior of the inflaton at the super-inflationary scale, have also been proposed, and it has been reported that these models yield a larger value of $n_\mathrm{s}$ compared to non-singular models~\cite{Kallosh:2025ijd,Ferreira:2025lrd}.
It would also be interesting to study the effects of quantum corrections in these models.
Furthermore, if the inflaton in $\alpha$-attractor models couples to heavy right-handed neutrinos or particles from Grand Unified Theories, such couplings should generally affect the inflaton potential through quantum corrections.
Building on these concrete examples, it would be phenomenologically interesting to delineate the parameter regions in which quantum corrections induce resonant enhancement of the GW signal, as discussed in this paper.

\section*{Acknowledgments}

We would like to thank Kaho Yoshimura for useful discussions.
K.N. thanks Shiladitya Porey for discussions regarding the results of recent CMB observations.
T.I. is supported by JST SPRING, Grant Number JPMJSP2108.
N.J.J. is grateful to the Office of the Dean of Academics
and the Department of Physics of Shiv Nadar Institution of Eminence for the kind financial support for a business trip to Japan in June 2025.
T.N. is supported by JSPS KAKENHI Grant No. JP26K07101 and JGC-Saneyoshi Scholarship
Foundation.
N.Y. is supported by JSPS KAKENHI Grant No.~JP21H05182.

\appendix

\section{CMB constraints}
\label{sec:CMB}

Over the past decade, the spectral index $n_\mathrm{s}$ and the tensor-to-scalar ratio $r$ have been constrained by numerous experiments, including Planck~\cite{Planck:2018vyg,Planck:2018jri}, Background Imaging of Cosmic Extragalactic Polarization~(BICEP)/Keck Array~\cite{BICEP:2021xfz}, the Atacama Cosmology Telescope~(ACT)~\cite{AtacamaCosmologyTelescope:2025blo,AtacamaCosmologyTelescope:2025nti}, and the South Pole Telescope~(SPT)~\cite{SPT-3G:2025bzu}.
Various combinations of the data obtained from these experiments have also been discussed, which involve taking into account the Wilkinson Microwave Anisotropy Probe~(WMAP) results~\cite{WMAP:2003elm}, and recent measurements of baryon acoustic oscillations~(BAO) by Dark Energy Spectroscopic Instrument~(DESI)~\cite{DESI:2024uvr,DESI:2024mwx,DESI:2025zpo,DESI:2025zgx}.
As thoroughly summarized in~\cite{Ellis:2025zrf}, $\sim 2\sigma$ (or more) tensions in central values of $n_\mathrm{s}$ have been reported in 2025 between the Planck results announced in 2018 and those by ACT and associated combinations.\footnote{For possible ways to address this deviation, see, e.g.,~\cite{Kallosh:2025rni,Dioguardi:2025vci,Gialamas:2025kef,Salvio:2025izr,Antoniadis:2025pfa,German:2025mzg,He:2025bli,Drees:2025ngb,Zharov:2025zjg,Haque:2025uri,Liu:2025qca,Gialamas:2025ofz,Byrnes:2025kit,Addazi:2025qra,Mondal:2025kur,Saini:2025jlc,Haque:2025uga,Hai:2025wvs,Heidarian:2025drk,Choudhury:2025vso,Pallis:2025gii,German:2025ide,Ellis:2025ieh,Ahmed:2025sfm,Aoki:2025ywt,Modak:2025bjv,Pallis:2025vxo,Ellis:2025bzi,Ahmed:2025eip,Fu:2025ciy,Alexandre:2025ixz,Addazi:2025agg,Pallis:2026cyz,Pallis:2026qci,McDonald:2026pbf,Solbi:2026muk,Ellis:2026ceb}.}
Following~\cite{Ellis:2025zrf}, we list three representative results:
\begin{itemize}
\item
$n_\mathrm{s} = 0.9649 \pm 0.0042\,(68\%\,\text{CL})$; the Planck result, including gravitational lensing, for the scalar tilt~\cite{Planck:2018vyg,Planck:2018jri}, announced in 2018.
In addition, the BICEP/Keck and BAO data~\cite{BICEP:2021xfz} can be incorporated to improve the constraints.
\item
$n_\mathrm{s} = 0.9752 \pm 0.0030\,(68\%\,\text{CL})$; the combination of the Planck data, including lensing, with the ACT and DESI BAO DR2 data~\cite{AtacamaCosmologyTelescope:2025blo}, reported in 2025 and referred to as ACT-P-DESI.
\item
$n_\mathrm{s} = 0.9684 \pm 0.0030\,(68\%\,\text{CL})$; the combined result of SPT, Planck, and ACT~\cite{SPT-3G:2025bzu}, announced in 2025, referred to as {SPT-P-ACT}.
\end{itemize}

\begin{figure}[t]
  \centering
  \includegraphics[width=0.48\textwidth]{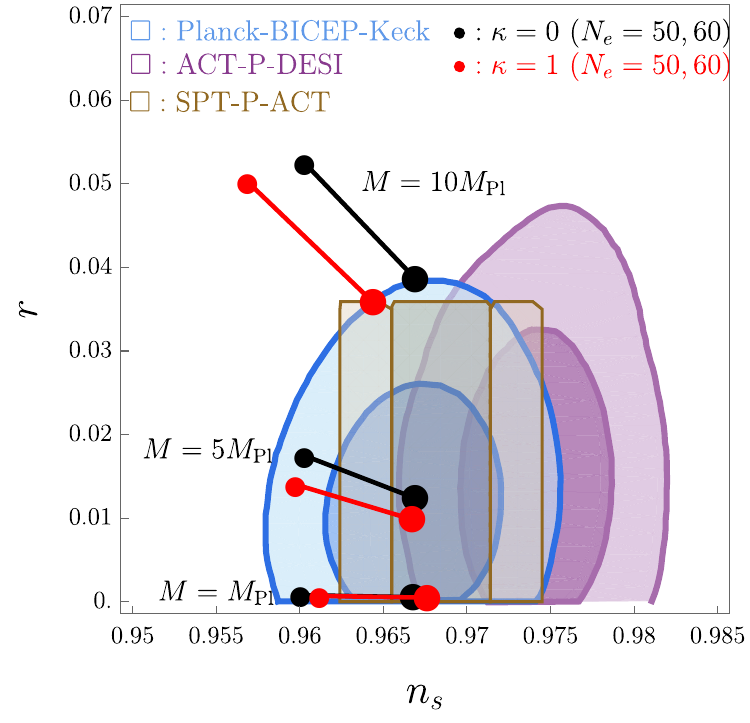} \quad
  \includegraphics[width=0.48\textwidth]{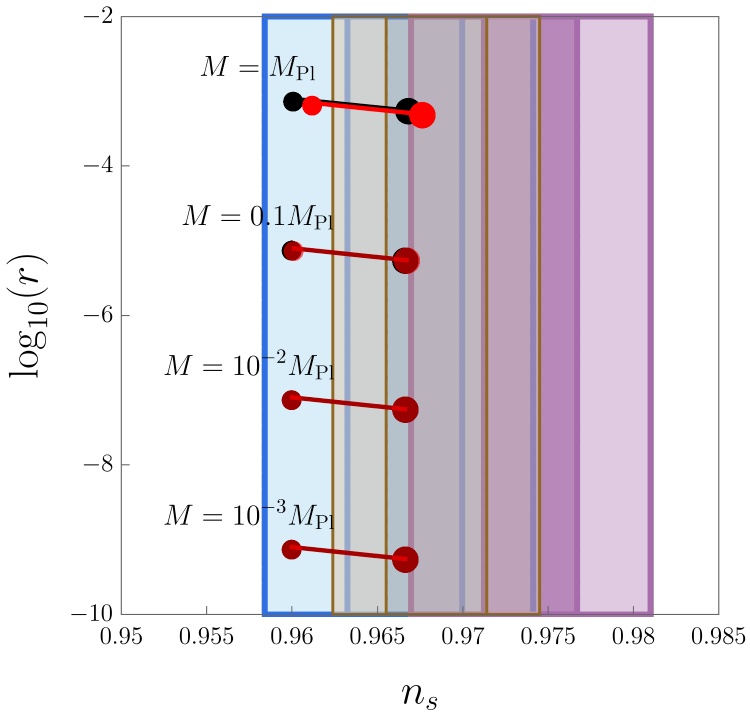}
\caption{
Constraints on the (quantum-corrected) $\alpha$-attractor T-model are shown with the $68\%$ and $95\%$ CL contours for the data combinations,
Planck/BICEP/Keck~\cite{BICEP:2021xfz} (blue),
ACT-P-DESI~\cite{AtacamaCosmologyTelescope:2025nti} (purple),
and SPT-P-ACT~\cite{SPT-3G:2025bzu} (brown).
For each $(M,\kappa)$ pair, the smaller and larger circles correspond to $N_e=50$ and $N_e=60$, respectively.
}
  \label{fig:CMB_ns-vs-r}
\end{figure}

The slow-roll conditions of the inflaton may also be expressed as conditions on the shape of the inflationary potential $V(\phi)$ in terms of the following parameters (see, e.g.,~\cite{Baumann:2009ds}),
\begin{align}
  \epsilon_{\mathrm{v}}(\phi)
  &:=
  \frac{\MPl^2}{2} \left(\frac{V_{,\phi}}{V} \right)^2,&
  \eta_{\mathrm{v}}(\phi)
  &:=
  \MPl^2 \frac{V_{,\phi\phi}}{V}.&
  \label{eq:def_slow-roll-parameters}
\end{align}
Note that, in the slow-roll regime
$\epsilon_{\mathrm{v}}, \, |\eta_{\mathrm{v}}| \ll 1$,
the background evolution is well approximated as
\begin{align*}
  H^2
  &\approx
  \frac{1}{3 \MPl^2} V(\phi) \approx \text{const.},&
  \dot{\phi}
  &\approx
  - \frac{V_{,\phi}}{3H},
\end{align*}
and the spacetime is approximately described as de Sitter space $a(t) \sim e^{H t}$.
Inflation ends when the slow-roll conditions are violated, which we define by
$\epsilon_{\mathrm{v}}(\phi_\mathrm{end}) \approx 1$,
where $\phi_\mathrm{end}$ denotes the inflaton field value at the end of inflation.
The $e$-folding number before inflation ends is defined as
\begin{align}
  N_e
  &:=
  \ln\left(\frac{a_\mathrm{end}}{a}\right)
  \approx
  \frac{1}{\MPl}
  \left|
  \int_{\phi_\mathrm{end}}^{\phi_\mathrm{CMB}}
  \frac{\dd{\phi}}{\sqrt{2\epsilon_{\mathrm{v}}(\phi)}}
  \right|\,,
  \label{eq:def_Ne}
\end{align}
where $\phi_\mathrm{CMB}$ represents the value of $\phi$ when the CMB scale exits the horizon.
The scalar spectral index $n_\mathrm{s}$ and the tensor-to-scalar ratio $r$ evaluated at $\phi = \phi_\mathrm{CMB}$ are
\begin{align}
  n_\mathrm{s}(\phi_\mathrm{CMB})
  &\approx
  1 + 2 \eta_\mathrm{v}(\phi_\mathrm{CMB}) - 6 \epsilon_\mathrm{v}(\phi_\mathrm{CMB}),&
  r(\phi_\mathrm{CMB})
  &\approx
  16 \epsilon_\mathrm{v}(\phi_\mathrm{CMB}).&
  \label{eq:def_CMB-observables}
\end{align}
The primordial scalar power spectrum ${\cal P}_s$ is parametrized as a power law,
\begin{align}
  {\cal P}_s(k)
  &=
  A_\mathrm{s} \left(\frac{k}{k_{\mathrm{piv}}}\right)^{ \left(n_\mathrm{s} - 1\right) + \frac{1}{ 2} n_{\mathrm{sk}} \ln\left(\frac{k}{k_{\mathrm{piv}}}\right) + \cdots },
\end{align}
where $A_\mathrm{s}$ is the scalar amplitude, $k_{\mathrm{piv}}$ is the pivot scale, and $n_{\mathrm{sk}} \coloneq \dd{n_\mathrm{s}}/\dd{(\ln{k})}$ is the running of the scalar spectral index $n_\mathrm{s}$.
This relation is useful for fixing $A_\mathrm{s}$ (see, e.g.,~\cite{Planck:2018jri,German:2020eyq}),
\begin{align}
  A_\mathrm{s}
  \approx
  \frac{1}{24\pi^2} \frac{V(\phi_\mathrm{CMB})}{ \epsilon_\mathrm{v}(\phi_\mathrm{CMB})} \frac{1}{\MPl^4}.
\end{align}
Note that the overall scale $\Lambda$ is determined through the amplitude of the scalar fluctuation spectrum as $A_\mathrm{s} \simeq 2.1 \times 10^{-9}$~\cite{Planck:2018vyg,Planck:2018jri}.
It is apparent from \eqref{eq:def_slow-roll-parameters} and \eqref{eq:def_CMB-observables} that $n_\mathrm{s}$ and $r$ do not depend on $\Lambda$.

\begin{figure}[t]
  \centering
  \includegraphics[width=0.48\textwidth]{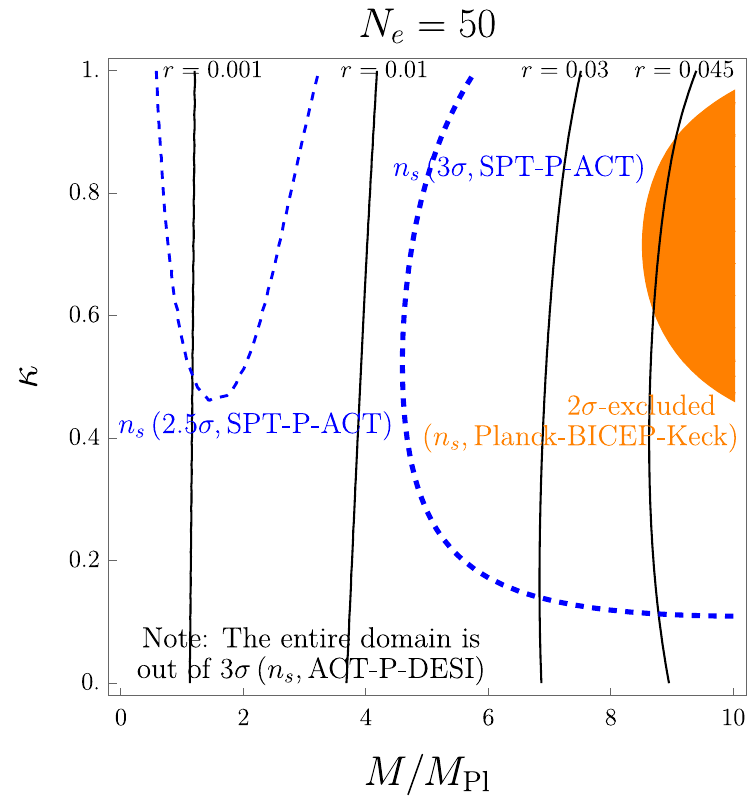} \quad
  \includegraphics[width=0.48\textwidth]{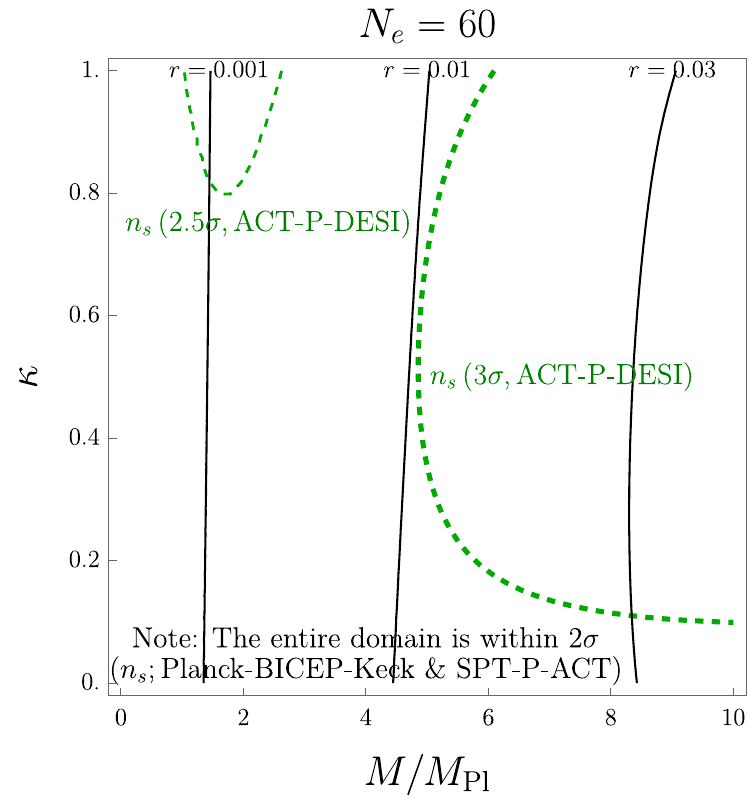}
\caption{
We summarize the CMB constraints on $n_\mathrm{s}$ and $r$ as functions of $M$ and $\kappa$, which are the free parameters of the quantum-corrected inflaton potential in \eqref{eq:Veff_full}.
The black contour lines indicate the corresponding values of $r$.
In each panel, the regions above the $2.5\sigma$ contour and to the left of the $3\sigma$ contour correspond to the parameter regions allowed within $2.5\sigma$ and $3\sigma$, respectively. For the observational constraints associated with the three values of $n_\mathrm{s}$, see the list at the beginning of this appendix.
}
\label{fig:CMB}
\end{figure}

Using \eqref{eq:def_slow-roll-parameters}, \eqref{eq:def_Ne} and \eqref{eq:def_CMB-observables}, we calculate $n_\mathrm{s}$ and $r$ for $N_e = 50$ and $60$ with the quantum-corrected inflation potential in \eqref{eq:Veff_full}; the results are summarized in Figs.~\ref{fig:CMB_ns-vs-r} and \ref{fig:CMB}.
For most of the $(M,\kappa)$ parameter regions under consideration, for both $N_e = 50$ and $60$, at least one of the three representative experimental results for $n_\mathrm{s}$ mentioned above lies within $2\sigma$.
It should be noted that the results are more consistent with the experimental data when $N_e = 60$.
When $N_e = 50$, there is a region that lies at least $2\sigma$ away from all three experimental results; however, this region is not relevant to the discussion of GWs in the main text.
We also comment on recent $95\%\,\text{CL}$ experimental upper bounds on $r$:
$< 0.036$ (Planck and BICEP/Keck)~\cite{BICEP:2021xfz}, $< 0.038$ (Planck, ACT, and BICEP/Keck)~\cite{AtacamaCosmologyTelescope:2025nti}, $< 0.25$ (SPT-3G)~\cite{SPT-3G:2025vtb}.
The first two of these impose restrictions on the parameter space when $M$ is large, but this aspect falls outside the scope of our discussion of GWs.
Here, the property that the effect of the corrections to $n_\mathrm{s}$ and $r$ is small when $\kappa$ is nonzero can be understood from the general shape of the (quantum-corrected) potential in the region where $\phi$ is large, as shown in Fig.~\ref{fig:potential_plot}.

\section{Details of the \texorpdfstring{{\CosmoLattice}}{CosmoLattice} simulations}
\label{sec:CosmoLattice}

This appendix summarizes the numerical setup used for the {\CosmoLattice} simulations in Sec.~\ref{sec:GW_simulation}.
We describe our choice of initial conditions, including the treatment of the initial fluctuation seeds, and examine the dependence of our results on the lattice resolution and IR cutoff, as well as their stability under further time evolution.

\subsection{Initial conditions}

\begin{figure}[t]
  \centering
  \includegraphics[width=0.6\textwidth]{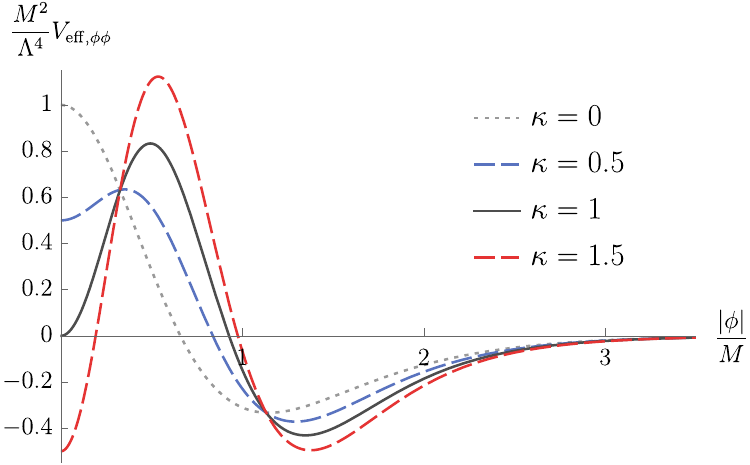}
  \caption{The second derivative of the effective potential \eqref{eq:Veff_full} for several values of $\kappa$, in units of $\Lambda^4/M^2$.
  For $\kappa=1$, the curvature is negative for $|\phi|>0.93M$ and asymptotically approaches zero at large field values.}
  \label{fig:potential_curvature}
\end{figure}

In {\CosmoLattice}, one has to specify the homogeneous field value $\phi_{\mathrm{i}}$ and its velocity $\dot{\phi}_{\mathrm{i}}$ as initial conditions of the simulation.
We determine $\dot{\phi}_{\mathrm{i}}$ by numerically evolving the background equation of motion together with the Friedmann equation from a region deep in the slow-roll regime, where the slow-roll approximation provides reliable initial conditions, up to $\phi=\phi_{\mathrm{i}}$.

\begin{table}[b]
  \centering
  \small
  \begin{tabular}{lccccc}
    \hline
    \rule{0pt}{2.8ex} & $M\,[\mathrm{GeV}]$ & $\Lambda\,[\mathrm{GeV}]$ & $\phi_{\mathrm{i}}\,[\mathrm{GeV}]$ & $\dot{\phi}_{\mathrm{i}}\,[\mathrm{GeV}^2]$ & $k_{\mathrm{IR}}\,[\mathrm{GeV}]$ \\
    \hline
    Figs.~\ref{fig:gw_M1}, \ref{fig:gw_M1_gt1} & $2.44\times10^{18}$ & $5.72\times10^{15}$ & $-0.9M=-2.19\times10^{18}$ & $1.21\times10^{31}$ & $1.84 \times 10^{12}$ \\
    Figs.~\ref{fig:gw_M0.1}, \ref{fig:gw_M0p1_gt1} & $2.44\times10^{17}$ & $1.86\times10^{15}$ & $-3M=-7.31\times10^{17}$ & $1.81\times10^{29}$ & $ 3.71 \times 10^{12}$\\
    Fig.~\ref{fig:gw_M0.01} & $2.44\times10^{16}$ & $5.90\times10^{14}$ & $-3M=-7.31\times10^{16}$ & $4.62\times10^{28}$ & $1.00 \times 10^{12}$ \\
    Fig.~\ref{fig:gw_smallinit} & $2.44\times10^{16}$ & $5.90\times10^{14}$ & $-0.5M=-1.22\times10^{16}$ & $0$ & $1.24 \times 10^{12}$ \\
    \hline
  \end{tabular}
  \caption{Potential parameters and initial conditions used in the lattice simulations for $\kappa=1$ in Sec.~\ref{sec:GW_simulation}. The values of $\Lambda$ are fixed by the CMB normalization for $N_e=60$.
  }
  \label{tab:simulation_parameters}
\end{table}

In principle, the subsequent dynamics should be insensitive to the choice of $\phi_{\mathrm{i}}$, provided that $|\phi_{\mathrm{i}}|$ is larger than the characteristic field value $\phi_*$ at which the oscillations begin.
Nevertheless, choosing an initial value with an excessively large magnitude is undesirable: it not only unnecessarily increases the simulation time, but may also introduce numerical errors because the characteristic field scale varies substantially during the evolution.
Moreover, as discussed in the following paragraph, for some choices of $\phi_{\mathrm{i}}$, the initial fluctuations cannot be generated appropriately.
Taking these considerations into account, we perform most of our simulations with either $\phi_{\mathrm{i}}=-0.9M$ or $\phi_{\mathrm{i}}=-3M$, except for the case shown in Fig.~\ref{fig:gw_smallinit}, for which we deliberately adopt artificial initial conditions.
The initial conditions used in the representative $\kappa=1$ simulations presented in Sec.~\ref{sec:GW_simulation}, together with the other model and simulation parameters, are summarized in Table~\ref{tab:simulation_parameters}.
In general, $\Lambda$ and $\dot{\phi}_{\mathrm{i}}$ depend on $\kappa$, but they can be determined in the same manner as described above.

\paragraph{Subtlety in initial fluctuations.}

A separate subtlety concerns the prescription used in {\CosmoLattice} to initialize the fluctuation seeds.
The initial seed modes are generated using an effective mass determined by the local curvature of the inflaton potential, $V_{\mathrm{eff},\phi\phi}(\phi_{\mathrm{i}})$, evaluated at the initial homogeneous field value~\cite{Figueroa:2021yhd}.
When this curvature is negative, the code neglects the negative mass-squared contribution and instead adopts the massless dispersion relation $\omega_k=|\mathbf{k}|$.
This prescription can potentially introduce a spurious dependence of the initial fluctuations on the choice of $\phi_{\mathrm{i}}$.

The curvature of the effective potential~\eqref{eq:Veff_full} for several values of $\kappa$ is shown in Fig.~\ref{fig:potential_curvature}.

As can be seen, the curvature becomes negative over part of the field range; specifically, for $\kappa=1$, it is negative for $|\phi| > 0.93M$.
The simulation should ideally be initialized at a field value for which the curvature is positive.
However, even when the curvature is negative, the resulting error in the initial fluctuation is expected to be small if the associated tachyonic scale
$\mu\coloneq\sqrt{|V_{\mathrm{eff},\phi\phi}(\phi_{\mathrm{i}})|}$
is much smaller than the momenta $k$ of the modes of interest.
As is also evident from Fig.~\ref{fig:potential_curvature}, for all values of $\kappa$ shown, the magnitude of the negative curvature becomes small at sufficiently large field values, such as $|\phi|=3M$.
For simulations with $\kappa\sim1$, we therefore choose $\phi_{\mathrm{i}}=-0.9M$ when the oscillation amplitude $\phi_*$ satisfies $\phi_*<0.9M$, and $\phi_{\mathrm{i}}=-3M$ when a larger initial value is needed to ensure $|\phi_{\mathrm{i}}|>\phi_*$.

\subsection{Cutoff dependence and late-time stability}

\begin{figure}[t]
  \centering
  \includegraphics[width=0.7\textwidth]{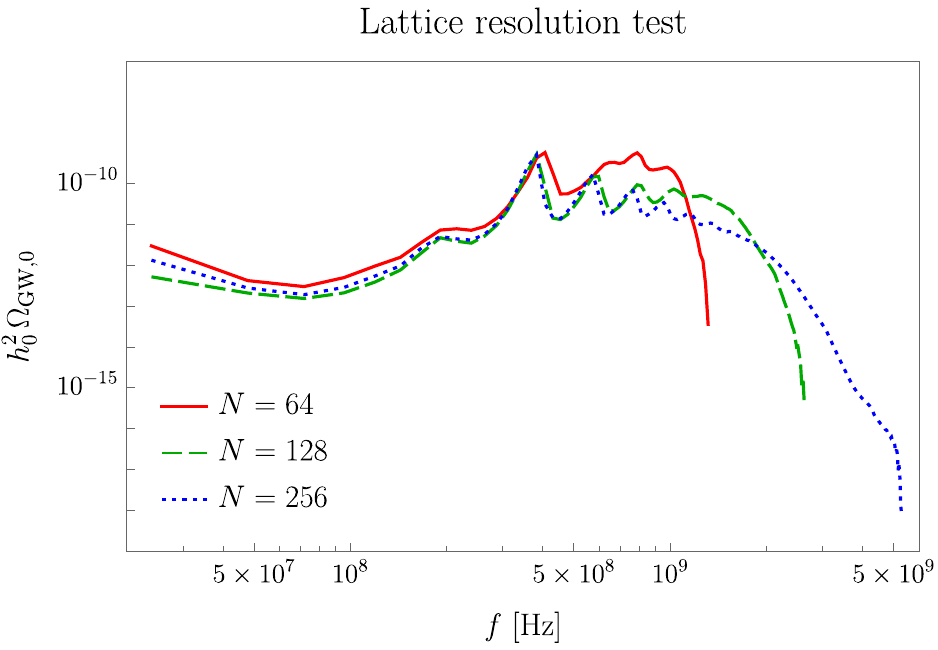}
  \caption{Convergence test with respect to the lattice resolution for the representative run with $M=\MPl$ and $\kappa=1$.
  The comparison between runs with different $N$ shows that the spectrum converges as the lattice resolution is increased, while the region nearest the ultraviolet cutoff is the most sensitive to finite-resolution effects.}
  \label{fig:conv_resolution}
\end{figure}

To demonstrate the numerical robustness of the simulations presented in Sec.~\ref{sec:GW_simulation}, we examine the dependence of the resulting GW spectra on the lattice cutoff scales and their stability under further time evolution.
Here, we present these checks only for representative cases, although we have confirmed that the other simulations in Sec.~\ref{sec:GW_simulation} exhibit the same robustness.

\paragraph{Lattice resolution.}

First, we compare the GW spectra obtained at different lattice resolutions, parameterized by the number of lattice points $N$, to identify a resolution at which all physically relevant scales are well separated from the UV cutoff $k_{\mathrm{UV}}$.
In Fig.~\ref{fig:conv_resolution}, we can see how the GW spectrum for the case with $M=\MPl$ and $\kappa=1$ approaches convergence as the number of lattice points $N$ is increased, with all other simulation parameters held fixed.
In particular, for smaller $N$, spurious features are visible near the UV cutoff, whereas these are substantially reduced for larger $N$.
Even for $N=256$, the spectrum very close to the cutoff should not be regarded as reliable.
However, the multiple peaks relevant for our discussion lie well inside the reliable region and already show good convergence at this resolution.

\begin{figure}[t]
    \centering
    \includegraphics[width=0.7\linewidth]{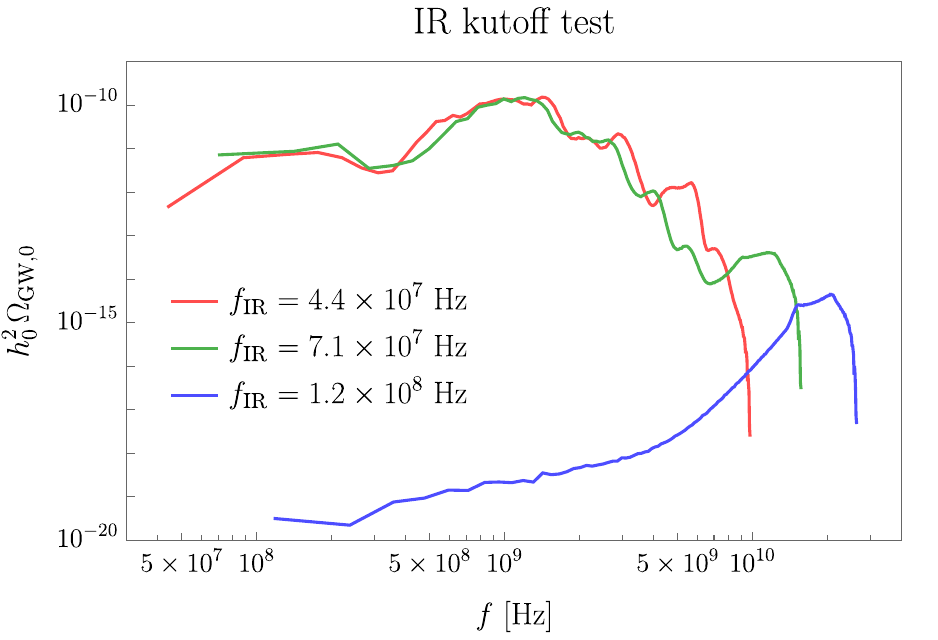}
    \caption{GW spectra at $M=\MPl$ and $\kappa=1.01$ for three values of the IR cutoff, with $N=256$. The red, green, and blue curves correspond to increasing values of $k_{\mathrm{IR}}$, respectively, with momenta converted into present-day frequencies. The red and green runs resolve the low-momentum modes responsible for the enhancement, whereas the blue spectrum is suppressed because $k_{\mathrm{IR}}$ exceeds the characteristic scale $k_*$ of the resonance.
}
\label{fig:conv_IR}
\end{figure}

\paragraph{IR cutoff.}

Next, we examine the dependence of the GW spectrum on the IR cutoff while keeping the number of lattice points fixed.
The results for $M=\MPl$ and $\kappa=1.01$ are shown in
Fig.~\ref{fig:conv_IR} (see also the right panel of
Fig.~\ref{fig:gw_M1_gt1}, which uses the IR cutoff listed in
Table~\ref{tab:simulation_parameters}).
Here, the IR cutoff momentum has been converted into the corresponding present-day cutoff frequency $f_{\mathrm{IR}}$, taking into account the redshift factor.
The red and green curves, corresponding to the two smaller IR cutoffs, agree closely over the intermediate-frequency range away from the cutoff.
In contrast, the spectrum with the largest IR cutoff, shown by the blue curve, differs drastically, with GW production strongly suppressed.

This behavior can be understood as a consequence of choosing an IR cutoff larger than the characteristic momentum scale $k_*$ associated with the relevant nonlinear dynamics, such that
\begin{equation}
k_* < k_{\mathrm{IR}}.
\end{equation}
Depending on the preheating mechanism, $k_*$ may correspond, for example, to the characteristic resonance scale $k_\mathrm{p}$ in quartic preheating or to the tachyonic scale $k_\mathrm{tach} \sim \sqrt{|V_{\mathrm{eff},\phi\phi}|}$ in tachyonic preheating.
When the above condition is satisfied, the lattice fails to capture the modes that drive the initial growth of fluctuations.

Importantly, even in this case, the spectrum may be affected at frequencies considerably higher than the IR cutoff $f_{\mathrm{IR}}$.
In general, fluctuations amplified around $k_*$ induce rescattering and nonlinear mode coupling, leading to the growth of fluctuations at other scales, typically including scales of order integer multiples of $k_*$, as discussed in Sec.~\ref{subsec:sim_M1}.
Indeed, Fig.~\ref{fig:conv_IR} shows that increasing the IR cutoff also removes the spectral feature around $10^9\,\mathrm{Hz}$.
This observation suggests that a step-like feature may arise as a secondary consequence of the growth of the IR modes.

\paragraph{Late-time stability.}

\begin{figure}[t]
  \centering
  \includegraphics[width=0.7\textwidth]{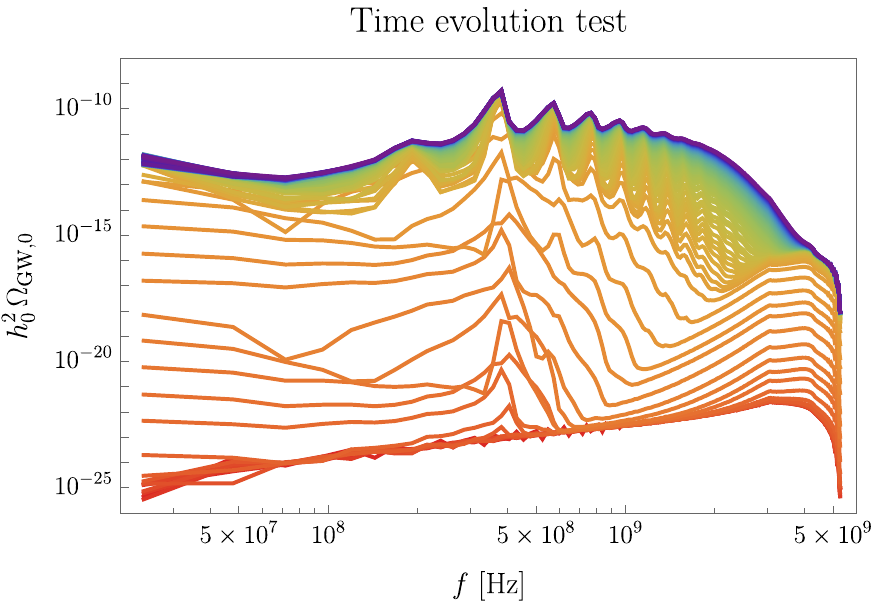}
  \caption{Convergence test based on the time evolution of the GW spectrum for the representative run with $M=\MPl$ and $\kappa=1$ at fixed $N=256$.
  The spectra are extracted at several times from the same simulation, with colors running from red at early times to purple near the end, to show that the spectrum has stabilized by the final stage.}
  \label{fig:conv_time}
\end{figure}

Finally, fixing the cutoff scales, we examine the time evolution of the GW spectrum for $M=\MPl$ and $\kappa=1$. The results are shown in Fig.~\ref{fig:conv_time}, with the colors shifting from red at the beginning of the simulation to purple toward the end.
The spectrum changes significantly at early times, reflecting the active growth stage of the source, but it converges during the final stage of the simulation.
This indicates that the simulation has been evolved long enough for the displayed final spectrum to be regarded as reliable.

{\small \bibliography{ref}}
\bibliographystyle{utphys}

\end{document}